\documentclass[review,3p]{elsarticle}

\usepackage{natbib}
\usepackage{array}
\usepackage{amsmath} 
\usepackage{amsthm}
\usepackage[mathlines,modulo]{lineno} 
\usepackage{enumitem}
\usepackage{xcolor}
\usepackage{hyperref}
\hypersetup{colorlinks=true,linkcolor=blue,citecolor=black}
\usepackage{soul}
\usepackage{booktabs}
\modulolinenumbers[5]

\usepackage{subcaption}
\usepackage{graphicx}

\journal{Transportation Research Part B}

\biboptions{authoryear}

\usepackage{algpseudocode}
\usepackage{makecell}

\begin{document}
	
	\begin{frontmatter}
		
		\title{\textbf{Optimal design of ride-pooling as on-demand feeder services}}
			\author[mymainaddress]{Wenbo Fan}
		\author[mysecondaryaddress]{Weihua Gu}
		\author[thirdaddress]{Meng Xu}
		
		\address[mymainaddress]{School of Transportation and Logistics, Southwest Jiaotong University; Email: wbfan@swjtu.edu.cn}
		\address[mysecondaryaddress]{Department of Electrical Engineering, The Hong Kong Polytechnic University; Email: weihua.gu@polyu.edu.hk (Corresponding author)}
		\address[thirdaddress]{School of Systems Science, Beijing Jiaotong University; Email: mengxu@bjtu.edu.cn}
		
		\begin{abstract}
			 The technology-enabled ride-pooling (RP) is designed as an on-demand feeder service to connect remote areas to transit terminals (or activity centers). We propose the so-called ``hold-dispatch'' operation strategy, which imposes a target number of shared rides (termed the ride-pooling size) for each vehicle to enhance RP's transportation efficiency. Analytical models are formulated at the planning level to estimate the costs of the RP operator and the patrons. Accordingly, the design problem is constructed to minimize the total system cost concerning the system layout (i.e., in terms of service zone partitioning), resource deployment (i.e., fleet size), and operational decisions (i.e., RP size). The proposed models admit spatial heterogeneity arising from the non-uniformity of demand distributions and service locations, and can furnish heterogeneous designs. Closed-form formulas for the optimal zoning and fleet size are developed, which unveil fundamental insights regarding the impacts of key operating factors (e.g., demand density and distance to the terminal). 
			 Extensive numerical experiments demonstrate (i) the effectiveness of heterogeneous service designs and (ii) the advantage of the proposed RP service with hold-dispatch strategy over alternative designs studied in the literature, i.e., RP with a ``quick-dispatch'' strategy and flexible-route transit, in a wide range of operating scenarios.  			 
			 These findings can assist transportation network companies and transit agencies in successfully integrating RP and transit services.
		\end{abstract}
		
		
	\end{frontmatter}
	
	\pagebreak
	
	\section{Introduction}
	
	\subsection{Research background and motivation}

	Advances in mobile internet and real-time location technologies are driving the rapid global growth of ride-sourcing mobility, with mixed impacts on urban transportation. On the one hand, travelers enjoy the convenience of ``door-to-door'' e-hailing services (e.g., Uber, Lyft, and Didi) without the expense of owning, driving, and parking private cars. On the other hand, the surge in on-demand personal travel has siphoned off patronage from public transportation and exacerbated traffic congestion \citep{graehler2019understanding,lee2019ride,nie2017can}. In order to mitigate the negative impacts and pursue sustainable mobility, the key to ride-sourcing mobility is to improve its transportation efficiency and coordinate its relationship with mass transit systems in a way that complements rather than competes with them. Many efforts have been made to connect the main form of ride-sourcing mobility---non-shared ride-hailing---as an on-demand feeder mode to transit networks \citep[e.g.,][]{aldaihani2004network,chen2017connecting,feigon2016shared,steiner2020strategic,stiglic2018enhancing,wen2018transit}.
	
	The emergence of shared/pooled ride-hailing, hereafter called ride-pooling (RP)\footnote{Also called ``ride-splitting'' in \cite{wang2019ridesourcing}.}, provides a promising solution for the above mission. Examples of RP services include Uber (Express) Pool, Lyft Line/Shuttle, Via Vanpooling, and Didi Express Pool, etc. Through RP, more than one patrons with close itineraries share a vehicle ride and split the trip cost. As a result of the increased vehicle occupancy and smart routing plans, ride-sourcing mobility’s transportation efficiency is improved compared to its non-shared counterpart. Positive externalities of RP include reduced vehicle distance traveled and lessened environmental pollution \citep{alonso2017demand,caulfield2009estimating,chan2012ridesharing,furuhata2013ridesharing,sperling2018three,yu2017environmental}. 
	
	The appealing features make RP more transit-friendly and integrable with conventional public transportation. 
	Using RP as an on-demand feeder to the transit, riders may benefit from the increased accessibility in terms of faster speed, direct connections to trip origins and destinations, prompt responses, affordable fees, etc. The transit system would profit from expanded service areas and increased ridership \citep{zhu2020analysis}. Additional systematic improvements can come from (re-)designing transit networks with larger-spaced lines and stations, which can increase the commercial speed of transit vehicles and reduce operating costs \citep{chen2017analysis,luo2019impact,liu2021mobility}. 
	
	Therefore, we aim to design RPs as feeder services and focus specifically on connecting remote areas (e.g., suburbs) to transit terminals (activity centers). Advanced models will be developed to account for RP’s key characteristics and operation regimes that are not fully explored in the literature. Next, we review the related studies to identify research gaps and summarize our contributions.

	\subsection{Related studies}
%
%
	The literature on RP is rich on operational research topics (e.g., vehicle-demand matching and fare splitting) but limited on optimal designs of RP as feeder services.\footnote{See \cite{wang2019ridesourcing,mourad2019survey,furuhata2013ridesharing,agatz2012optimization} for good reviews. Modeling RP differs significantly from modeling non-shared taxis and mass transit (either on-demand or fixed-route). The former must jointly consider demand pooling, demand-vehicle matching, vehicle dispatching, and vehicle routing \citep{daganzo2019public}. Thus, the existing taxi-type or transit design models, {such that in \cite{aldaihani2004network,nourbakhsh2012structured},}  cannot be directly used to analyze RP services.} Two works closely related to this study are discussed below in detail.   
	
	The first work is \cite{zhu2020analysis}, which modeled RP as both a feeder and an alternative to transit and examined the impact of operation strategies (regarding fare discounts and pooling promotions) on an existing public transit system. Through sensitivity analysis, they found that a win-win situation could be achieved by operating RP as a feeder service to the trunk transit. This feeder-trunk structure can increase the number of RP orders for transportation network companies (TNCs), boost transit ridership for the government, and provide low-cost services for passengers. 
	
	Nonetheless, the study of \cite{zhu2020analysis} is limited in several ways. First, it did not optimize the RP design (e.g., the number of shared rides served per vehicle). Second, the RP operator's cost was ignored and important operational decisions (e.g., the fleet size of RP vehicles) were treated as given. Third, the study ignored the zoning strategy of RP, i.e., partitioning the entire service region into several zones to further improve the operational efficiency and reduce the system-wide cost. Zoning has an important impact on limiting the detour time of RP vehicles, which was not explicitly modeled in the referenced work.
	
	The second relevant work, \cite{liu2021mobility}, addressed some of the above limitations. {The study examined the RP service under a so-called ``quick-dispatch'' (QD) strategy (the service is thus termed the RP-QD service), where a vehicle is dispatched immediately to pick up the first-come request and continues to receive new requests until that first patron is onboard or a predefined number of requests (two as specified in the paper) have been received.} An optimization model was proposed for minimizing both the RP operator's cost and patrons' trip times to determine the operational features (e.g., the fleet size) and the physical layout (e.g., the RP service region size) of an integrated RP and transit network.  
	The authors employed an overarching analytical framework to derive closed-form formulas for system metrics (e.g., the RP operator's cost and patrons' trip time components). They found that the integrated system can significantly outperform the conventional single-mode transit system and the one fed by fixed-route buses. They also verified that an optimally-designed transit system fed by RP always dominates the one fed by non-shared ride-hailing. 

 Nevertheless, for the sake of tractability, \cite{liu2021mobility} is based on several idealized assumptions and simplifications regarding the demand and layout of transit systems, which limit the practical value of their models. First, it assumes that patrons' origins and destinations are uniformly distributed throughout the entire region of study. This assumption does not reflect the general reality. Especially for feeder transport systems, the demand densities to and from a terminal usually {vary} with the distance to the terminal \citep{fan2018joint,zhen2023feeder}. Second, the study stipulates that the design variables, such as fleet size and zone size of RP services, are spatially uniform. {Their designs can be further improved} even under the uniform demand distribution, because the optimal service design in a local area is affected by the trip distances of patrons in that area. It is clear that transporting the same amount of demand farther away requires more fleet. Thus, the deployment and operation of transportation resources should vary with the location \citep{su2019heterogeneous}. {Third, it did not consider further partitioning of a transit terminal's service region, which is reasonable and potentially beneficial for large suburbs and satellite cities. And lastly, the number of shared rides per vehicle is set to be no more than two.} The choice of this parameter is critical to RP's operations and can affect its performance, as opposed to alternative feeder modes, such as non-shared taxis and flexible-route transit. {Note that the design features related to the latter two points (i.e., the service region partitioning and the number of rides per vehicle) can both be spatially heterogeneous.} In short, the referenced work lacks the consideration of spatial heterogeneity and its fundamental impact on the feeder system designs.
	
	It is worth noting that hybrid systems of RP and transit have also been evaluated in several simulation studies \citep[e.g.,][]{gurumurthy2020first,pinto2020joint,ma2017spatial}, which also reported promising results in terms of the improved system performance. These simulation models, however, lack analytical tools to guide the system design.
	
	There are also plenty of studies on other feeder modes, such as shared bikes and car-pooling (which require patrons to be able to ride bikes or drive cars) \citep[e.g.,][]{wu2020optimal,luo2021joint}, non-shared ride-hailing (which is subject to low transportation efficiency and high fees) \citep{aldaihani2004network,zhu2020analysis}, and flexible-route transit (FlexRT, also called paratransit, demand-adaptive transit, or microtransit) \citep{chang1991optimization,quadrifoglio2009methodology}. FlexRT is another type of on-demand mobility service. {It differs from RP in that FlexRT (as appeared in most works in the literature, e.g., \cite{chang1991optimization,chen2022optimized, kim2019optimal}) dispatches vehicles one by one at a fixed headway, while in RP multiple vehicles are distributed in a network, accepting requests simultaneously. Thus, in FlexRT each vehicle needs to traverse the entire service zone to pick up the patrons, while in RP a vehicle only accepts the closest requests to its current location.}\footnote{It is necessary to delineate the two concepts, which are usually confused in the literature. Their inherent operational differences lead to divergent system performance, as will soon be demonstrated in this paper.} Interested readers are referred to \cite{sangveraphunsiri2022jitney,kim2019optimal,luo2019impact,chen2017analysis,nourbakhsh2012structured} for recent efforts in FlexRT designs. Comparisons between the performances of RP and FlexRT will be made in Section \ref{sec_numerics}.

	 \subsection{Research gaps and our contributions}
	 In light of the above inadequacies in the literature, we intend to fill the research gap in RP-based feeder services by considering arbitrary demand distributions and furnishing optimal heterogeneous designs. Specifically, we are concerned about four macro-level questions: 
	 \begin{enumerate}[label=(\roman*)]
	 	\item Is partitioning an RP feeder service region into local zones beneficial, and what is the best partition?
	 	\item How many RP vehicles are needed, and how to deploy them among the zones? 
	 	\item How many rides should an RP vehicle in a specific zone serve (i.e., what is the optimal ride-pooling size)? 
	 	\item Under what conditions would the RP-based feeder service outperform FlexRT? 
	 \end{enumerate}
	 
	 Answers to these questions are synthetically affected by many factors, such as the level and distribution of demand, patrons' value of time (VOT), the operation regime and cost of RP vehicles, and the location and size of the service region under study. To find the answers, we propose new analytical models for optimal designs of RP-based feeder systems under spatially-heterogeneous demand, which simultaneously consider service zone partitions, RP vehicle deployment, and operational strategies.	
   
	 There have been two commonly-used methods to address spatial heterogeneity in transit network design problems. The first method is building discrete mathematical programming (MP) models on graph networks with demand given in the form of an origin-destination (OD) matrix \citep{DuranMiccoVansteenwegen2022,ibarra2015planning,kepaptsoglou2009transit}. These MP models, however, contain numerous variables and parameters and often fall into the category of NP-hard problems, which suffer from the dilemma between solution efficiency and quality \citep{magnanti1984network, Schobel2012}. The other method is the so-called ``continuum/continuous approximation'' (CA) approach, which expresses variables (e.g., demand density, service zone size, and line spacing) as continuous functions of spatial coordinates \citep[e.g.,][]{badia2014competitive,chen2015optimal,ouyang2014continuum}. {Ideally, these two approaches should be complementary \citep{daganzo2019public,daganzo1987increasing,hall1986discrete}. On the one hand, discrete MP is adept at precisely evaluating particular problems while considering realistic constraints. On the other, CA facilitates the creation of parsimonious models that can more clearly demonstrate the fundamental interrelations between key variables. The latter can be utilized to identify which specific problems deserve the efforts required for accurate computations \citep{newell1971dispatching}.} 
	 
	 The consideration of spatial heterogeneity introduces more challenges in the model formulation and solution compared to developing homogeneous optimal designs under uniform demands. The literature on CA-based transit network designs holds some clues on how to overcome these challenges. For instance, by exploiting the decomposition techniques, closed-form solutions or optimal analytical conditions may be derived thanks to the parsimony of these CA models \citep{wirasinghe1981spacing,luo2021joint,mei2021planning,medina2013model}. {Hence, in this paper we will develop CA models that can assist transit agencies and TNCs with planning-level decisions on RP service designs.} 

	The main contributions of this paper are fourfold:
	\begin{enumerate}[label=(\roman*),itemjoin=\\]
		\item It is the first paper to integrate spatial heterogeneity into the design of RP-based feeder systems, of which parsimonious models are formulated to minimize costs for both the operator and patrons. The models concern three location-dependent decision variables, i.e., zone size, fleet density, and ride-pooling size. By exploiting the local decomposition property of the CA formulation, closed-form solutions are obtained for the zone size and fleet density as functions of location, from which new fundamental insights are drawn and discussed. Built upon the above analytical results, optimal ride-pooling sizes are found numerically with high efficiency. 
		\item The importance and benefit of spatially heterogeneous RP service designs are revealed. For instance, under optimized designs, the zone size, fleet size, and ride-pooling size all tend to vary with distance to the connected transit terminal, even when demand is uniformly distributed. {Sizeable cost savings are observed compared to uniform designs. Furthermore, the zoning strategy can reduce system costs by up to 12\% under a highly-heterogeneous demand. These findings have valuable implications for the practical implementation of RP-based feeder services.}
		\item A new operation scheme, termed the ``hold-dispatch'' (HD) strategy, is proposed for RP services, resulting in what is referred to as the RP-HD service. This strategy holds vehicles until receiving a target number of requests and dispatches them for pickups following well-planned Traveling Salesman Problem (TSP) tours. This strategy is appealing because the optimized ride-pooling size (i.e., the number of requests) can guarantee transportation efficiency. Numerical comparisons with its counterpart, i.e., the RP-QD service strategy \citep{daganzo2019public,liu2021mobility}, show the superiority of our proposed strategy in terms of significant system cost savings (up to 8\%) over a wide range of scenarios.
		\item Comparisons are also made with the conventional flexible-route service, FlexRT (see Section \ref{sec_numerics} for detailed descriptions). The results show that our RP-HD service consistently dominates FlexRT in all tested scenarios. Notable system cost savings (up to 6\%) are achieved in scenarios with low to medium demand levels and more heterogeneous demand distributions. However, the advantage of RP-HD gradually diminishes as the demand level increases because FlexRT exhibits a greater degree of economies of scale (EOC) (i.e., a greater marginal benefit from increased demand) than RP-HD. 
	\end{enumerate}

	The remainder of this paper is organized as follows: The next section formulates models for RP as feeder services, followed by numerical tests in Section \ref{sec_testt}. In particular, Section \ref{sec_numerics} conducts experiments demonstrating the performance of the proposed system as compared to its two counterparts, i.e., RP-QD and FlexRT. The last section concludes the paper and discusses future research directions.
	
	\section{Modeling ride-pooling as on-demand feeder service}
	We first present the basic concepts and assumptions of RP-based feeder services in Section \ref{sec_concept} and then describe the system's steady state in Section \ref{sec_steady_state}. Next, Sections \ref{sec_operator_cost} and \ref{sec_patron_trip_time} build models for the RP operator's cost and patrons' trip time, respectively. The optimal design problem is formulated and {the analytical properties of its optimal solution} are developed in Section \ref{sec_optimal_design}. For readers' convenience, Table \ref{tab_notations} in \ref{notation} summarizes the notations used in the paper.
	
	\subsection{Basic concepts and assumptions} \label{sec_concept}
\begin{figure}[h]
		\centering
		\includegraphics[width=0.6\linewidth]{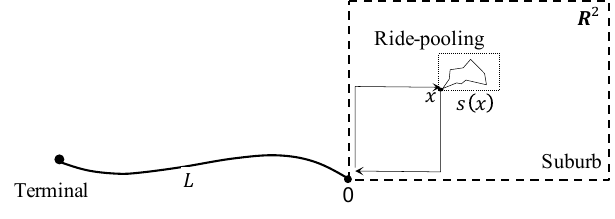}
		\caption{Layout of the ride-pooling  feeder service.}
		\label{fig.layout}
	\end{figure}

	To isolate the effect of spatial heterogeneity on RP-based feeder designs, we consider only the feeder trip segment without including the backbone transit. The system layout is shown in Figure \ref{fig.layout}, where a suburban region is connected to a transit terminal by an $ L $ km-long freeway. Similar layouts have been adopted in previous studies, such as \cite{chen2022optimized,liu2020effects,kim2019optimal,kim2014integration,chang1991optimization}.
	{We denote $ \mathbf{R} $ as the suburban region and a two-dimensional vector $ x \in \mathbf{R} $ as an arbitrary location in the region; $ x=(0,0) $ indicates the freeway entrance.} 
	
	 A TNC or transit agency operates a fleet of e-hailing vehicles with dedicated drivers to serve patrons traveling {outbound (from their residences in the suburb to the terminal) and inbound (from the terminal to the suburb). The study period is a typical morning peak when the outbound demand dominates. (Evening peak periods can be studied in similar ways.)} Suppose the RP operator divides the suburb into multiple local service zones and seeks to determine the zone size, denoted by $ s(x) $ [km$ ^2 $/zone], the fleet density, $ f(x) $ [vehicles/km$ ^2 $], and the outbound ride-pooling size (i.e., occupancy), $ u(x) $ [patrons/vehicle], which can vary with location $ x \in \mathbf{R} $.\footnote{{We will explain momentarily why the inbound ride-pooling size is not a design variable in our paper.}}
    {Note that these design variables are location-dependent, representing a heterogeneous system design. The advantage of heterogeneous designs over the conventional homogeneous designs studied in the literature will be demonstrated in Section \ref{sec_testt}, even under a uniform demand.}
	
	To facilitate model formulation, we adopt the following assumptions that were also commonly made in the literature \citep[e.g.,][]{liu2021mobility,daganzo2019general,kim2019optimal,clarens1975operating}.
	\begin{enumerate}[label=(\roman*)]
		\item  The RP operation is under the government's regulation aiming at minimizing the total system cost, consisting of the operator's and patrons' costs.
		\item The demand per hour for RP is exogenously given and fixed. The origins of the outbound trips and the destinations of the inbound trips are temporally uniform and spatially slow-varying\footnote{This ``slowly-varying demand'' assumption had been adopted in the literature of CA models of optimal transit design to ensure no large approximation errors in the estimation of locally decomposable metrics. Further discussion on the robustness of CA models with sharp changes in demand density can be found in \cite{hall1986discrete,daganzo1987increasing}; and Chapter 3 of \cite{daganzo2005logistics}.}, which can be represented by continuous density functions $ \lambda_u(x) $ and $ \lambda_v(x),  x \in \mathbf{R}  $ [trips/km$ ^2 $/h], respectively. Here the subscripts $ u $ and $ v $ denote outbound and inbound trips, respectively. We further assume {$ \lambda_u(x) > \lambda_v(x), x \in \mathbf{R} $}, reflecting the tidal demand pattern in a typical morning peak.  
		\item The RP operator employs a proximity-based algorithm that assembles the outbound trip requests and matches them to the closest available vehicles within each service zone. A matched vehicle is dispatched to pick up the outbound patrons and deliver them to the transit terminal. (However, other than the QD strategy assumed in the literature, we assume that the vehicle will be dispatched only after collecting all the target trip requests; see the last paragraph of this section for more details.) 
		{The same vehicle will pick up the inbound patrons waiting at the terminal (if any) and deliver them back to the same service zone.} Such a round trip constitutes an operation cycle. {Note that due to the dominance of outbound demand in the morning peak, the vehicle will not wait for inbound patrons at the terminal; i.e., it will return to its origin zone immediately even without a passenger.} It is further assumed that a vehicle picks up or drops off one patron per stop.
	\end{enumerate}

	{Since the demand is given and the vehicle flows in the outbound and inbound directions are equal, $u(x)$ and the (average) inbound ride-pooling size, $v(x)$, must satisfy {the following relationship in an undersaturated system (which will be guaranteed in the following optimal designs)}:} 
    {
        \begin{linenomath*}
		\begin{equation}       \frac{u(x)}     {v(x)}=\frac{\lambda_u(x)}{\lambda_v(x)}.		 \label{eq_v}
		\end{equation}
	\end{linenomath*}}
{Thus, the inbound ride-pooling size is not a design variable in the proposed model.}
 
	For outbound vehicles, we propose a ``hold-dispatch'' (HD) strategy. Under the HD strategy, each outbound vehicle is held to wait for the pooling of $u(x)$ requests and {is dispatched only when all the $u(x)$ requests are matched}.  After it is dispatched, the vehicle follows a route plan according to the optimal TSP tour and does not accept any new requests. Upon arrival at the terminal, the vehicle drops off the $u(x)$ outbound patrons, picks up any inbound patrons waiting at the terminal, and immediately starts delivering them back to the service zone according to the optimal TSP tour. Thus, each vehicle’s tour consists of two separate legs: {collection and delivery}.\footnote{There are other routing strategies, e.g., those that mix the drop-offs and pick-ups according to the proximity or some alternating rules \citep{daganzo1978approximate}. We leave these options for future study.} 
	
	\subsection{The system's steady state} \label{sec_steady_state}
	We are interested in the optimal service design for the system in steady states. We describe the states of the RP vehicles using the notation adopted mainly from \cite{daganzo2019general}: The $ n_{ij} $ indicates the number of vehicles in state $ (i,j) $, where $ i $ is the number of patrons in a vehicle and $ j $ the number of requests assigned to it {that are waiting} to be picked up. {For example, $(0,0)$ indicates an idle vehicle; $(0,u)$ a vehicle with all $u$ outbound requests assigned; $(u,0)$ a vehicle that has picked up all the $u$ requests; and $(v,0)$ a vehicle that has delivered the outbound patrons to the terminal, picked up the inbound ones, and is ready to return to the service zone.} 
	
\begin{figure}[h]
		\centering
		\includegraphics[width=0.4\linewidth]{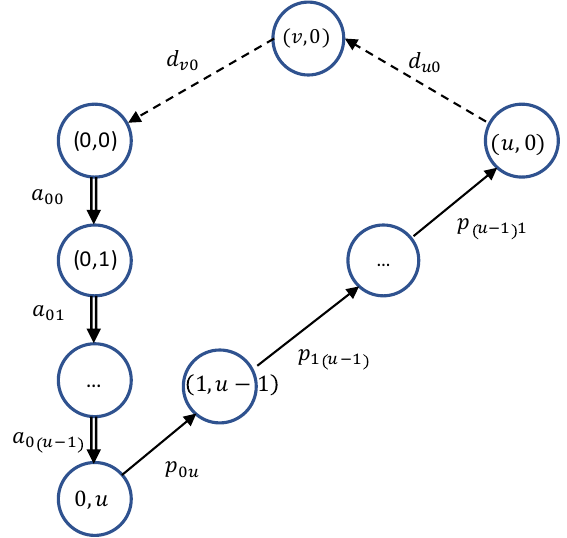}
		\caption{Workload transition network for vehicles.}
		\label{fig_workload}
	\end{figure}

	The workload state transitions of these vehicles can be visualized in Figure \ref{fig_workload}. {The nodes and links represent the workload states and the transitions between them, respectively.} The double-solid, single-solid, and dashed arrows represent the request assignments, (outbound) pick-ups, and (outbound and inbound) deliveries, respectively. The $ \left\{a_{ij},p_{ij},d_{u0}, d_{v0} \right\} $ are the corresponding transition flow rates [vehicles/h], {where $d_{u0}, d_{v0}$ differentiate between outbound and inbound deliveries.} {The vehicle numbers and flow rates can be functions of the location} $ (x) $ (e.g., $ n_{ij}(x) $).
	
	In the steady state, the flow rates in Figure \ref{fig_workload} satisfy
	\begin{linenomath*}
	\begin{equation}
		a_{00} = ... = a_{0(u-1)}= p_{0u}=...=p_{(u-1)1} = d_{u0} = d_{v0}.
        \label{flow_rates}
	\end{equation}
	\end{linenomath*}

    {From the given demand rate, we have:}
    \begin{linenomath*}
		\begin{equation}
				a_{0j}(x) = \frac{\lambda_{u}(x)s(x) }{u(x)}, j\in \left\{0,...,u(x)-1\right\}, \label{liitle_law_a}
		\end{equation}
	\end{linenomath*}
    where $ \frac{1}{u(x)} $ in Eq. (\ref{liitle_law_a}) yields the probability of requests being assigned to an available vehicle in {one of the} $ u(x) $ states, {i.e., $(0,0)$, $(0,1)$, ..., $(0,u(x)-1)$}.

	According to Little’s law \citep{little1961proof}, the transition flow rates of vehicle states in the local service zone at $ x \in \mathbf{R} $ should have the following relationships:
	\begin{linenomath*}
		\begin{subequations}
			\begin{align}
				p_{ij}(x) &= \frac{n_{ij}(x)}{t_p(x)}, i = u (x)- j, j\in \left\{1,...,u(x)\right\},\label{liitle_law_b}\\
				d_{u0}(x) &= \frac{n_{u0}(x)}{t^d_u(x)}, \label{liitle_law_c}\\
				d_{v0}(x) & = \frac{n_{v0}(x)}{t^d_v(x)}, \label{liitle_law_d}
			\end{align}
		\end{subequations}
	\end{linenomath*}
	
	The $ t_p(x) $ [hour/patron] in Eq. (\ref{liitle_law_b}) is the average travel time for a vehicle to pick up one patron. Under the proposed HD strategy, $ t_p (x)$ can be approximated by the average travel time per pick-up in the TSP tour,
	\begin{linenomath*}
		\begin{subequations} \label{t_p}
			\begin{align}
				t_p (x) &= \frac{1}{u(x)} \frac{\ell_u(x)}{V'} ,  \\
				\ell_u (x) &=k \sqrt{\frac{u(x)}{f_0(x)}}, \label{tsp_u}
			\end{align}
		\end{subequations}
	\end{linenomath*}
	where $ \ell_u (x) $ [km] is the total TSP tour length for picking up $ u(x) $ outbound patrons\footnote{{ The formula, in the general form of $ D = k \sqrt{N\cdot A}$ (where $D$ denotes the expected TSP tour length, $N$ is the number of visiting points, and $A$ is the area of the service zone), were derived for both closed and open TSP tours with different values of $k$ \citep{beardwood1959shortest,stein1978asymptotic,stein1978scheduling,daganzo1984length} (see footnote 1 in \cite{stein1978asymptotic} in particular). The two types of TSP differ in that the former requires the vehicle to return to the starting point (i.e., the first point of $N$), whereas the latter does not. In this study, the RP vehicle needs not to return to the first drop-off/pick-up point.}}, and $ V' $ [km/h] is the vehicles' average operating speed, discounted from the non-stop operating speed $ V $ to account for pick-up and drop-off delays \citep{kim2019optimal}.\footnote{{ Alternatively, one can construct $V'(x) = 1/(1/V + u(x) \tau)$ (with $\tau$ being the average delay per stop) for particular zones at $x \in \mathbf{R}$. This remedy will not alter the mathematical property of the following optimal design problem. For the sake of comparability, we remain using a constant $V'$ as in the literature \citep{chang1991optimization,kim2019optimal}. }} The $ k $ is a dimensionless constant related to the network topology. The $ \frac{1}{f_0(x)} \equiv  \frac{s(x)}{\sum_{j=0}^{u(x)-1} {n_{0j}(x)}} $ [km$ ^2 $/vehicle] is the vehicle's average service area containing $ u(x) $ closest requests, and $ f_0(x) $ [vehicles/km$ ^2 $] indicates the density of the available vehicles accepting new requests. Although Eq. (\ref{tsp_u}) is an approximation for non-uniform demands, our simulation tests show that the error is generally lower than 3\%, which becomes negligible in the scale of the system cost.
	
	The $ t^d_u(x) $ [hour] in Eq. (\ref{liitle_law_c}) is the average line-haul travel time from the service zone at $x$ to the terminal. {Since the vehicles' last stopping points vary in the neighborhood of $x$,} it can be approximated by
	\begin{linenomath*}
		\begin{equation}  
			t^d_u (x)= \frac{ \left( \|x\| + L \right) }{V}. \label{t^d_u}
		\end{equation}
	\end{linenomath*}

	The $ t^d_v(x) $ [hour] is the average delivery time of an inbound vehicle from the terminal to the service zone, which is the sum of the line-haul travel time and the TSP-based delivery tour time:
	\begin{linenomath*}
		\begin{subequations}  \label{t^d_v}
			\begin{align}
				t^d_v(x) &= \frac{ \left( \|x\| + L \right) }{V} + \frac{ \ell_v(x)}{V'}, \\
				\ell_v(x) &= k \sqrt{ v(x) s(x)},
                \label{l_v}
			\end{align}
		\end{subequations}
	\end{linenomath*}
	where $ \ell_v(x) $ [km] is the total TSP tour length for delivering $ v(x) $ inbound patrons within a zone of size $ s(x) $. {The $ v(x) $ is obtained from Eq. (\ref{eq_v}).}\footnote{The $v(x)$ satisfying Eq. (\ref{eq_v}) may take a fractional value. Thus, Eq. (\ref{l_v}) should be viewed as an approximation. However, its error only has a modest impact on the optimal system design since the inbound demand is small in the morning peak.}
	
	Next, we use the above steady-state relationships to derive the operator's cost and patrons' trip time. 
	
	\subsection{Operator cost} \label{sec_operator_cost}
	The RP operator's cost depends on the {fleet size}, which is derived as follows. 
	
	Combining Eqs. (\ref{flow_rates}, \ref{liitle_law_a}, \ref{liitle_law_b}, \ref{t_p}) yields
	\begin{linenomath*}
		\begin{align}
			n_{ij}(x) & = \frac{\lambda_{u}(x) s(x)}{u(x)}  t_p(x) =\frac{\lambda_{u}(x) s(x)}{u(x)} \frac{k}{V'}\sqrt{\frac{1}{u(x) f_{0}(x)}}, i=u(x)-j,j \in \left\{ 1,...,u(x)\right\}.
		\end{align}
	\end{linenomath*}

 	Similarly, combining Eqs. (\ref{eq_v}, \ref{flow_rates}, \ref{liitle_law_a}, \ref{liitle_law_c}, \ref{liitle_law_d}, \ref{t^d_u}, \ref{t^d_v}) gives
	\begin{linenomath*}
		\begin{equation}
			n_{u0}(x)+n_{v0}(x)=\frac{\lambda_{u}(x) s(x) }{u(x)}\left[\frac{2\left(\|x\|+L\right)}{V}+ \frac{\sqrt{\lambda_v(x)}}{\sqrt{\lambda_{u}(x)}}\frac{k \sqrt{u(x)s(x)}}{V'}\right].
		\end{equation}
	\end{linenomath*}
	
	Thus, the density of the operating fleet in the service zone at $  x $ is expressed as
	\begin{linenomath*}
		\begin{align}
			\notag
			f(x) =  & f_0 (x)+\frac{ \sum_{j=1}^{u(x)} {n_{(u(x)-j)j}(x)} }{s(x)} + \frac{n_{u0}(x) + n_{v0}(x)}{s(x)}, \\
			= & f_{0}(x)+\frac{\lambda_{u}(x)k}{V'}\sqrt{\frac{1}{u(x)f_{0}(x)}}+\frac{\lambda_{u}(x)}{u(x)}\left[\frac{2\left(\|x\|+L\right)}{V}+\frac{\sqrt{\lambda_v(x)}}{\sqrt{\lambda_{u}(x)}}\frac{k \sqrt{u(x)s(x)}}{V'}\right].
		\end{align}
	\end{linenomath*}
	And the total {operating fleet size} in the suburban region is
	\begin{linenomath*}
		\begin{equation}
			F =  \int_{x\in \mathbf{R}} { f(x) dx}.
		\end{equation}
	\end{linenomath*}

	Following previous studies \citep[e.g.,][]{chang1991optimization,kim2014integration,kim2019optimal}, the operator’s total cost, $ U_a $ [\$/h], is estimated by
	\begin{linenomath*}
		\begin{align} \label{U_a}
			\notag
			U_a = & \pi_f F, \\
			= & \pi_{f}\int_{x\in\mathbf{R}}{\left[f_{0}(x)+\frac{\lambda_{u}(x)k}{V'}\sqrt{\frac{1}{u(x)f_{0}(x)}}+\frac{\lambda_{u}(x)}{u(x)}\left(\frac{2\left(\|x\|+L\right)}{V}+\frac{\sqrt{\lambda_v(x)}}{\sqrt{\lambda_{u}(x)}}\frac{k \sqrt{u(x)s(x)}}{V'}\right)\right]dx},
		\end{align}
	\end{linenomath*}
	where $ \pi_f $ (\$/vehicle-h) is the average operating cost per vehicle hour, synthetically accounting for (i) the amortized capital cost (related to, e.g., vehicle purchase, rent, and maintenance) in the operation hour and (ii) the operational cost (e.g., drivers’ wage and fuel cost). 
	
	
	\subsection{Patron trip time} \label{sec_patron_trip_time}
	{The following two subsections model the trip time} for outbound and inbound patrons, respectively. 
	\subsubsection{Outbound patrons' trip time}
	The average trip time per outbound patron in a zone at $ x $ consists of three components: (i) the waiting time at home for pooling, i.e., the waiting time before the assigned vehicle begins the collection, denoted by $ t^w_u (x) $, (ii) the time (either at home or in a vehicle) during the vehicle's pickup tour, denoted by $ t^r_u (x) $; and (iii) the line-haul travel time from the last pickup location to the transit terminal, $ t^d_u (x) $. They are formulated as follows. 
	
	(i) Under the HD dispatching strategy, $ t^w_u (x) $ satisfies
	\begin{linenomath*}
		\begin{align}
			t^w_u (x) = \frac{\sum_{j=1}^{u(x)-1}jn_{0j}(x)}{\lambda_u\left(x\right)s\left(x\right)} = \frac{u(x)\left(u(x)-1\right)}{2}\frac{f_0(x)s(x)/u(x)}{\lambda_{u}\left(x\right)s\left(x\right)} = \frac{\left(u(x)-1\right)}{2}\frac{ f_{0}(x)}{\lambda_u\left(x\right)}.
		\end{align}
	\end{linenomath*}
    {This is because $n_{0j}(x)=\frac{f_0(x)s(x)}{u(x)}$, $\forall j \in \left\{1,...,u(x)-1 \right\}$.}

	(ii) The $ t^r_u (x)$ can be estimated by the optimal TSP tour travel time as
	\begin{linenomath*}
		\begin{equation}
			t^r_u (x)= \frac{k}{V'}\sqrt{\frac{u(x)}{f_{0}(x)}}.
		\end{equation}
	\end{linenomath*}
	
	(iii) The $ t^d_u(x) $ is given in Eq. (\ref{t^d_u}).
	
	Therefore, the total trip time of all outbound patrons is
	\begin{linenomath*}
		\begin{align}
			\notag
			U_u = & \int_{x \in \mathbf{R}} { \lambda_{u}(x) \left[t^w_u(x) + t^r_u(x) + t^d_u(x) \right] dx}, \\
			= & \int_{x\in\mathbf{R}}{\left[\frac{\left(u(x)-1\right)}{2}f_{0}(x)+\frac{\lambda_{u}(x)k}{V'}\sqrt{\frac{u(x)}{f_{0}(x)}}+\lambda_{u}(x)\frac{\|x\|+L}{V}\right]dx}.
		\end{align}
	\end{linenomath*}
	
	\subsubsection{Inbound patrons' trip time}
	{Under the zoning strategy, vehicles are assigned to specific zones. Inbound patrons must wait to board and travel with the vehicles heading to their desired zone.} Specifically, inbound patrons destined for the zone at $ x $ are subject to: (i) the terminal waiting time for vehicle arrivals, $ t^w_v(x) $; (ii) the line-haul travel time between the terminal and the first drop-off point in the zone, $ t^s_v(x) $; and (iii) the in-vehicle travel time in the TSP-based delivery tour, $ t^r_v(x) $. 
	
	(i) The $ t^w_r(x) $ is half of the vehicles' arrival headway{, $\frac{1}{d_{u0}(x)}$. From Eqs. (\ref{flow_rates}, \ref{liitle_law_c}, \ref{t^d_u}), we have:}
	\begin{linenomath*}
		\begin{equation}
			{t^w_v(x) = \frac{1}{2}\frac{\|x\| + L}{V n_{u0}(x)} = \frac{1}{2} \frac{u(x)}{\lambda_{u}(x)s(x)}} . 
		\end{equation}
	\end{linenomath*}
	
	(ii) The $ t^s_v(x) $ is {identical to $ t^d_u(x) $,} i.e.,
	\begin{linenomath*}
		\begin{equation}
			t^s_v(x) = \frac{\|x\| + L}{V}.
		\end{equation}
	\end{linenomath*}
	
	(iii) The $ t^r_v(x) $ is obtained by half of the TSP tour time,
	\begin{linenomath*}
		\begin{equation}
			t^r_v(x) = \frac{1}{2} \frac{\ell_v(x)}{V'} = \frac{k \sqrt{v(x)s(x)}}{2V'}  = \frac{\sqrt{\lambda_v(x)}}{\sqrt{\lambda_{u}(x)}}\frac{k \sqrt{u(x)s(x)}}{2V'}.	
		\end{equation}
	\end{linenomath*}
	
	Thus, the total trip time of all inbound patrons is
	\begin{linenomath*}
		\begin{align}
			\notag
			U_v &=  \int_{x \in \mathbf{R}} { \lambda_{v}(x) \left[t^w_v(x) +  t^s_v(x)  +  t^r_v(x)\right] dx}\\
			& = \int_{x\in\mathbf{R}}{\left[\frac{1}{2}  \frac{\lambda_{v}(x) u(x)}{\lambda_{u}(x)s(x)} + \lambda_{v}(x)\left(\frac{\left(\|x\|+L\right)}{V}+\frac{\sqrt{\lambda_v(x)}}{\sqrt{\lambda_{u}(x)}}\frac{k \sqrt{u(x)s(x)}}{2V'}\right)\right]dx}.
		\end{align}
	\end{linenomath*}
	
	\subsection{Optimal design} \label{sec_optimal_design}
	The generalized system cost, denoted by $ Z $, is a weighted sum of the operator's cost and the patrons' trip time, as given by
	\begin{linenomath*}
		\begin{align}
			Z = \frac{U_a}{\mu} + U_u + U_v,
		\end{align}
	\end{linenomath*}
	where $ \mu $ is the patrons' average VOT and is used to convert the monetary cost to the time unit. 
	
	The optimal design problem is constructed to minimize the generalized system cost concerning $ s(x), f_0(x), u(x) $, as expressed by
	\begin{linenomath*}
		\begin{subequations} \label{design_problem}
			\begin{align}
				\min_{s(x),f_0(x),u(x)} &Z = \int_{ x \in \mathbf{R} } { \left[z_1 (x,s(x),u(x)) + z_2 (x,f_0(x),u(x)) + z_3(x,u(x))\right] dx}, \label{obj}
			\end{align}
		\text{subject to:} 
		\begin{align}
			s(x),f_{0}(x)>0, u(x) \in \left\{1,2,..., C\right\}, \forall x\in \mathbf{R}, 
		\end{align}
		where the integrands in (\ref{obj}) are local system cost components at location $ x $, as given by
		\begin{align}
			&z_1(x,s(x),u(x)) =  \frac{\lambda_{u}\left(x\right)}{u(x)} \frac{\sqrt{\lambda_{v}(x)}}{\sqrt{\lambda_{u}(x)}} \frac{\frac{\pi_{f}}{\mu}k\sqrt{u(x)s(x)}}{V'}+\frac{1}{2}\left[\frac{\lambda_{v}(x) u(x)}{\lambda_{u}(x)s(x)}+\lambda_{v}(x)\frac{\sqrt{\lambda_{v}(x)}}{\sqrt{\lambda_{u}(x)}}\frac{k\sqrt{u(x)s(x)}}{V'}\right] ,\\
			&z_2 (x,f_0(x),u(x)) = \frac{\pi_{f}}{\mu}\left[f_{0}\left(x\right)+\frac{\lambda_{u}(x)k}{V'}\sqrt{\frac{1}{u(x)f_{0}\left(x\right)}}\right]
			+\frac{\left(u(x)-1\right)f_{0}\left(x\right)}{2}+\frac{\lambda_{u}\left(x\right)k}{V'}\sqrt{\frac{u(x)}{f_{0}\left(x\right)}}, \\
			&z_3(x,u(x)) = \frac{\lambda_{u}\left(x\right)}{u(x)}\frac{\left(2\frac{\pi_{f}}{\mu}+u(x)\right)\left(\|x\|+L\right)}{V}+\lambda_{v}(x)\frac{\left(\|x\|+L\right)}{V}, 
		\end{align}
		\end{subequations}
	\end{linenomath*}
	and $ C $ is the vehicle capacity [seats/vehicle].
	
	Note in (\ref{design_problem}), {which ensures the system is undersaturated}, that the {design variable $ u(x) $ is integer-valued and identical for all vehicles serving the same zone, reflecting a clear (and hard) target.}\footnote{Alternatively, one may adopt a ``soft'' target based on certain pooling criteria, e.g., a given upper bound on the waiting time for pooling. We leave the investigation of this modeling choice to future research. {Further details can be found in the discussion on jitney services} in \cite{daganzo2019public}.} On the other hand, the mean occupancy $ v(x) $ of inbound vehicles is a continuous variable because the inbound vehicles are dispatched without holding. The actual occupancy of an inbound vehicle varies around the mean $ v(x) $.
	
	{To solve the mathematical program (\ref{design_problem}), we first note that if $ u(x) $ is fixed at the optimal $ u^*(x), x \in	\mathbf{R} $, then by the calculus of variations, (\ref{design_problem}) can be decomposed into two independent subproblems at each $ x \in \mathbf{R} $:}
	
	(i) minimizing {$ z_1(x,s(x)|u^*(x))\equiv  z_1(x,s(x), u(x)=u^*(x)) $} concerning $ s(x) $, i.e., 
	\begin{linenomath*}
		\begin{subequations}
			\begin{align}
				\min_{s(x)} z_1(x,s(x)|u^*(x)) =  \frac{\lambda_{u}\left(x\right)}{u^*(x)} \frac{\sqrt{\lambda_{v}(x)}}{\sqrt{\lambda_{u}(x)}} \frac{\frac{\pi_{f}}{\mu}k\sqrt{u^*(x)s(x)}}{V'}+\frac{1}{2}\left[\frac{\lambda_{v}(x) u^*(x)}{\lambda_{u}(x)s(x)}+\lambda_{v}(x)\frac{\sqrt{\lambda_{v}(x)}}{\sqrt{\lambda_{u}(x)}}\frac{k\sqrt{u^*(x)s(x)}}{V'}\right] ,
			\end{align}
		\text{subject to:} 
			\begin{align}
			s(x) >0, 
			\end{align}
		\end{subequations}
	\end{linenomath*}
and (ii) minimizing {$ z_2(x,f_0(x)|u^*(x))\equiv z_2(x,f_0(x),u(x)=u^*(x)) $} concerning $ f_0(x) $, i.e.,
\begin{linenomath*}
	\begin{subequations}
		\begin{align}
			\min_{f_0(x)} z_2&(x,f_{0}(x)|u^*(x))= \frac{\pi_{f}}{\mu}\left[f_{0}\left(x\right)+\frac{\lambda_{u}(x)k}{V'}\sqrt{\frac{1}{u^*(x)f_{0}\left(x\right)}}\right]
			+\frac{\left(u^*(x)-1\right)f_{0}\left(x\right)}{2}+\frac{\lambda_{u}\left(x\right)k}{V'}\sqrt{\frac{u^*(x)}{f_{0}\left(x\right)}}, 
		\end{align}
	\text{subject to:}
		\begin{align}
			f_{0}(x)>0, 
		\end{align}
	\end{subequations}
\end{linenomath*}

We note that $ z_1(x,s(x)) $ and $ z_2(x,f_0(x)) $ are posynomial functions regarding $ s(x) $ and $ f_0(x) $, respectively. A posynomial is a sum of monomials, each of which has the form of $Cx_1^{\alpha_1}x_2^{\alpha_2}...x_n^{\alpha_n}$ where $C\geq0, x_i\geq0$, and $\alpha_i$ are real numbers. A posynomial function can be converted to a convex function using a technique called geometric programming \citep{boyd2004convex}. Thus, the two subproblems yield unique global optima. Specifically, the optima, $s^*(x|u^*(x))$ and $f_0^*(x|u^*(x))$, are given in Sections \ref{opt_zone} and \ref{opt_fleet}, respectively. Finally, Section \ref{opt_rp} presents the method for finding $u^*(x)$ and some analytical insights.  

\subsubsection{Optimal zone size}
\label{opt_zone}
The $ z_1(x,s(x)) $ is in the form of $ A\cdot s(x)^{\frac{1}{2}}+B\cdot s(x)^{-1} $ where $A$ and $B$ are positive. Thus, the first-order condition of subproblem (i) yields the global optimum:
\begin{linenomath*}
	\begin{equation} \label{sol_s}
		s^{*}(x|u^*(x))=\left[\frac{2V' \mu\sqrt{\lambda_{v}(x)}}{k\left(2\pi_{f}+\mu\lambda_{v}(x)\frac{u^*(x)}{\lambda_{u}\left(x\right)}\right)}\right]^{\frac{2}{3}}\frac{u^*(x)}{\lambda_{u}\left(x\right)}.
	\end{equation}
\end{linenomath*}

	\subsubsection{Optimal fleet density}
    \label{opt_fleet}
	The $ z_2(x,f_0(x)) $ is in the form of $A\cdot f_0(x)+B\cdot f_0(x)^{-\frac{1}{2}}$ where $ A, B>0 $. Therefore, $f^*_0(x)$ can be derived from the first-order condition of subproblem (ii):
	\begin{linenomath*}
		\begin{equation}
	\label{sol_f0}		f_{0}^{*}\left(x|u^*(x)\right)=\left[\frac{k\left(\pi_{f}+\mu u^*(x)\right)\lambda_{u}\left(x\right)}{\left(2\pi_{f}+\mu \left(u^*(x)-1\right)\right)V' \sqrt{u^*(x)}}\right]^{\frac{2}{3}},
		\end{equation}
	\end{linenomath*}
	{Note that $f^*_0(x)$ is not directly related to the inbound demand. This is not surprising since $f^*_0(x)$ is the number of available vehicles accepting outbound trip requests.}
	
	With the above results, the optimal fleet density in any zone $ x \in \mathbf{R} $ is
	\begin{linenomath*}
		\begin{align} \label{sol_f}
			\notag
			f^{*}(x|u^*(x))= & \left[\frac{k\lambda_{u}(x)}{V' \sqrt{u^*(x)}}\right]^{\frac{2}{3}}\frac{3\pi_{f}+\mu\left(2u^*(x)-1\right)}{\left[2\pi_{f}+\mu\left(u^*(x)-1\right)\right]^{\frac{2}{3}}\left[\pi_{f}+\mu u^*(x)\right]^{\frac{1}{3}}} \\ &+ 2\frac{\lambda_{u}(x)}{u^*(x)V}\left(\|x\|+L\right)+\left(\frac{k\lambda_{v}(x)}{V'}\right)^{\frac{2}{3}}\left(\frac{2\mu}{2\pi_{f}+\mu\lambda_{v}(x)\frac{u^*(x)}{\lambda_{u}\left(x\right)}}\right)^{\frac{1}{3}}.
		\end{align} 
	\end{linenomath*}

\subsubsection{Optimal ride-pooling size {and some insights}}
\label{opt_rp}

	To find the optimal ride-pooling size, $ u^*(x) $, plugging (\ref{sol_s}) and (\ref{sol_f0}) into the original problem (\ref{design_problem}) yields the problem with a single unknown variable $ u^*(x) $. Although there is no closed-form solution, the finite-valued $ u^*(x) $ can be {easily found via enumeration}. The process is quite straightforward since the vehicle capacity $ C $ is usually small, e.g., $ C=4 $ for sedans and $ C=6 $ for vans. {Specifically, we solve the following program by enumeration for each $ x \in \mathbf{R} $:}
	\begin{linenomath*}
		\begin{subequations}
			\begin{align}
			u^*(x) = \text{argmin}_{u(x) \in \left\{1, 2,...,C\right\}} \left\{z_u(x,u(x)) {+ z_v(x,u(x))}+  z_0(x)\right\}, 
			\end{align}
		\text{where,} 
		\begin{align}
			z_{u}(x,u(x))
			&=\frac{\pi_{f}}{\mu}\left(\frac{k\lambda_{u}(x)}{V' \sqrt{u(x)}}\right)^{\frac{2}{3}}\frac{3\pi_{f}+\mu\left(2u(x)-1\right)}{\left[2\pi_{f}+\mu\left(u(x)-1\right)\right]^{\frac{2}{3}}\left[\pi_{f}+\mu u(x)\right]^{\frac{1}{3}}} \notag \\
			&+\left(\frac{k\lambda_{u}(x)}{V' \sqrt{u(x)}}\right)^{\frac{2}{3}}\left[\frac{\left(\pi_{f}+\mu u(x)\right)}{\left(2\pi_{f}+\mu\left(u(x)-1\right)\right)}\right]^{\frac{2}{3}}\frac{\left[u(x)-1\right]}{2} \notag \\
			&+\left(\frac{k\lambda_{u}\left(x\right)u(x)}{V'}\right)^{\frac{2}{3}}\left[\frac{\left(2\pi_{f}+\mu\left(u(x)-1\right)\right)}{\left(\pi_{f}+\mu u(x)\right)}\right]^{\frac{1}{3}}+\frac{\lambda_{u}\left(x\right)}{u(x)}\frac{\pi_{f}}{\mu}\frac{2\left(\|x\|+L\right)}{V} , \label{z_u} \\
			{z_{v}(x,u(x))} &=\left(\frac{k\lambda_{v}(x)}{2V'}\right)^{\frac{2}{3}}\frac{6\pi_{f}+3\mu\frac{\lambda_{v}(x)}{\lambda_{u}(x)}u(x)}{2\left(2\mu^{2}\pi_{f}+\mu^{3}\frac{\lambda_{v}(x)}{\lambda_{u}(x)}u(x)\right)^{\frac{1}{3}}},\\
			z_0(x)  &= \frac{\left(\lambda_{u}\left(x\right)+\lambda_{v}(x)\right)\left(\|x\|+L\right)}{V}. \label{z_0}
		\end{align}
		\end{subequations}
	\end{linenomath*}
	
	The above analytical results unveil several fundamental insights regarding the RP-HD strategy: 
\begin{itemize}
 	\item[(i)] Eq. (\ref{sol_s}) implies that the inbound demand is the very reason for {optimal zone partitioning.} To see it, consider that if the inbound demand is negligible, such that $ \lambda_{v}(x) \approx 0 $, then the original problem will be reduced to subproblem (ii) because $ z_{1}(x)\approx 0 $. This implies that we can freely choose the zone size $ s(x) $. We can also deduce that zone sizes have a secondary impact on the system performance when the inbound demand is much smaller than the outbound one (e.g., a tidal commute demand in morning rush hours).  This is expected, because there are multiple ($f_0(x)s(x)$) vehicles scattered around a service zone, accepting outbound trip requests simultaneously. Each vehicle serves its neighborhood only. {This kind of ``distributed service" acts as if a service zone is further divided into ``subzones." As a result, the effect of optimizing $s(x)$ is diluted. The ``subzoning" effect is the main reason for the difference between RP-HD and designs without such a hierarchical zoning structure (e.g., the FlexRT service discussed in Section \ref{sec_numerics} and \ref{sec_FlexRB}).} 
  
  On the other hand, for a significant $ \lambda_{v}(x) $, both the operational variable $ u(x) $ and the outbound demand $ \lambda_{u}(x) $ affect the optimal zone size $ s^*(x) $. 
	\item[(ii)] Eq. (\ref{sol_f}) implies that (a) {the fleet size is a linear function of the line-haul travel distance}; and (b) vehicles' line-haul speed $ V $ (with a power of $ -1 $) has a more significant effect on reducing the fleet size and the associated costs than the average operating speed in the service zones, $ V' $ (which has a power of $ -\frac{2}{3} $). 
	\item[(iii)] The results speak to the necessity of the spatially heterogeneous designs of the ride-pooling sizing and zoning strategies. As seen in Eq. (\ref{z_u}), {the coefficients of $ u(x) $ involve terms related to $ \lambda_{u}(x) $ and $ \|x\| $.} It implies that the optimal $ u^*(x) $ depends on {two spatially heterogeneous factors: the outbound demand density and the service zone location. Thus, $ u^*(x) $ should be spatially heterogeneous, even if the demand ($ \lambda_u(x) $) is spatially uniform. Moreover, the spatial heterogeneity in $ u^*(x) $ leads to the spatial heterogeneity in $ s^*(x) $ and $ f^*(x) $.} 
	
	In contrast, if $ u(x) $ is specified as a constant for all $x \in \mathbf{R}$, then $ s^*(x) $ would be invariant in space under the uniform demand, while $ f^*(x) $ would increase linearly with $ \|x\| $.
\end{itemize}

%
	

	\section{Numerical studies} \label{sec_testt}
	
	\subsection{Experimental set-up} \label{set-up}
\begin{figure}[!htbp]
		\centering
		\begin{subfigure}{0.45\textwidth}
			\centering
						\includegraphics[width = \textwidth]{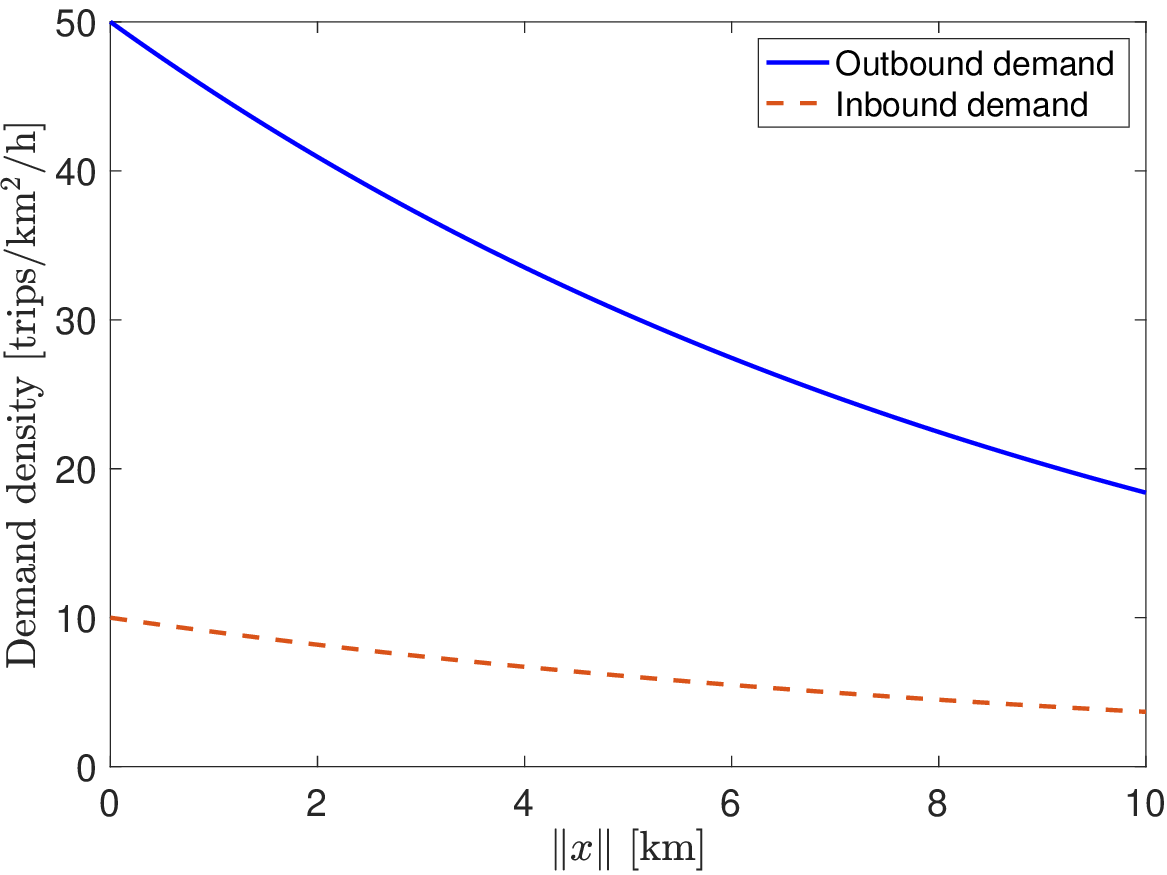}
			\caption{Less-heterogeneous demand ($\eta_u , \eta_v=0.1$)}
		\end{subfigure}
		\begin{subfigure}{0.45\textwidth}
						\includegraphics[width = \textwidth]{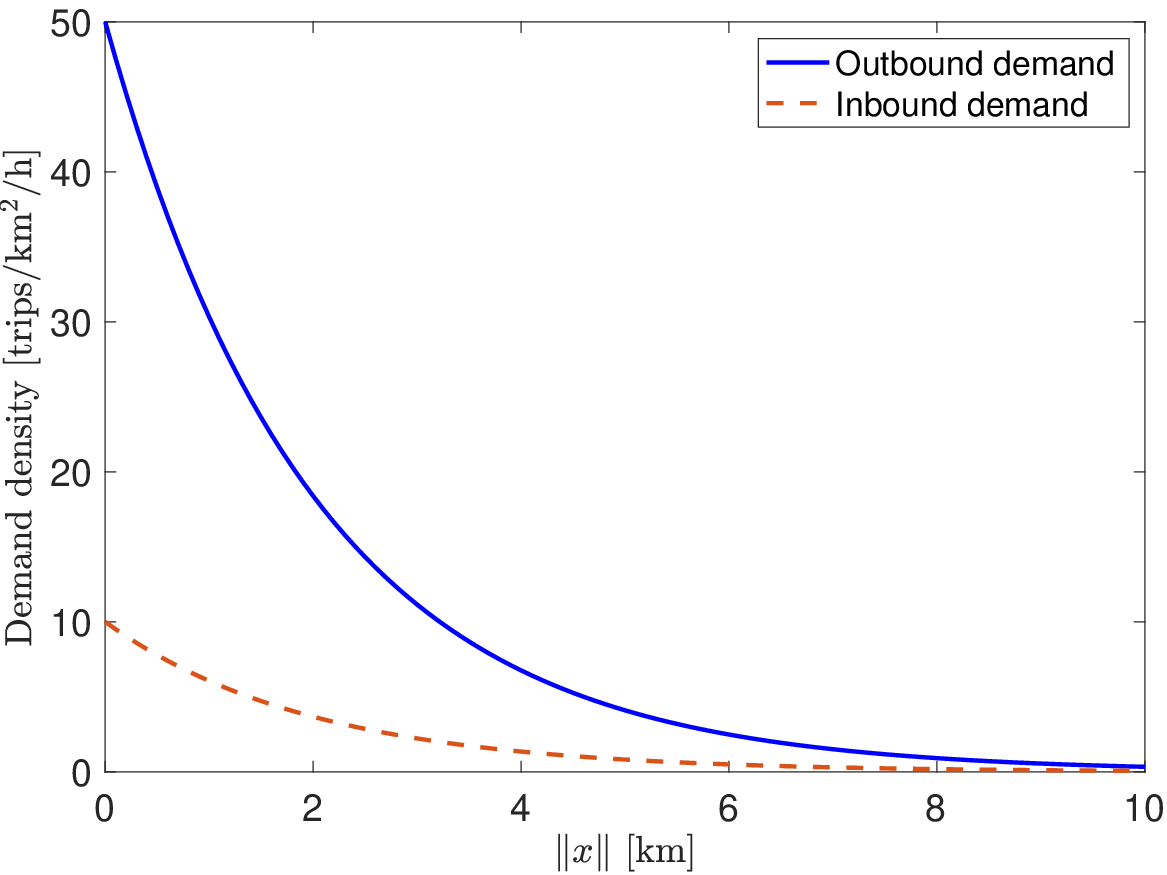}
			\caption{Highly-heterogeneous demand ($\eta_u , \eta_v=0.5$)}
		\end{subfigure}
		\caption{Demand patterns.}
		\label{demand_patterns}
	\end{figure}

	For demonstration, we consider a square-shaped suburban area with a length and width of 5 km and a distance-decaying demand distribution \citep{fan2018joint,luo2019paired,li2012modeling}, 
	\begin{linenomath*}
		\begin{align} \label{demand_fun}
			&\lambda_u(x) = \bar{\lambda}_u \text{exp}(-\eta_u \|x\|), & \lambda_v(x) = \bar{\lambda}_v \text{exp}(-\eta_v \|x\|), x\in \mathbf{R},
		\end{align}
	\end{linenomath*}
	where $ \bar{\lambda}_u = 50,\bar{\lambda}_v =10 $ [trips/km$ ^2 $/h], and the parameters $ \eta_u, \eta_v \in \left\{0,0.1,0.5\right\} $ reflect the sensitivity of demand to the distance, as shown in Figure \ref{demand_patterns}. A special case of (\ref{demand_fun}) is the uniform demand distribution when $ \eta_u $ and $ \eta_v $ take 0 and $ \lambda_u(x)=\bar{\lambda}_u $ and $ \lambda_v(x)=\bar{\lambda}_v, \forall x \in \mathbf{R}$.

	Other model parameters’ values are given in Table \ref{tab_param_value}. They are adopted from previous studies \citep[e.g.,][]{kim2019optimal}. The distance between the suburb and the terminal is set as $ L=5 $ km. {For the sake of comparability, our computations are done in the Manhattan metric, and take $ k=1.15 $ from the literature \citep{chang1991optimization,kim2014integration,kim2019optimal}, which are based on Stein's open-tour TSP formula \citep{stein1978asymptotic}.}

	\begin{table}[h] 
		\centering
		\caption{Parameter baseline values}
			\begin{tabular} {|c|c|c|c|}
				\hline 
				\textbf{Parameter} & \textbf{Baseline value} & \textbf{Parameter} & \textbf{Baseline value}\\
				\hline 
				$ V, V' $ & 30, 25  [km/h] & $ \mu $ & 25 [\$/h] \\
				\hline
				$ \pi_{f} $& 5.9 [\$/vehicle-h]& $ C $ & 4 [seats/vehicle] \\
				\hline
		\end{tabular}
	\label{tab_param_value}
	\end{table}

	{The accuracy of the proposed CA models is verified through simulation experiments conducted under uniform, less-, and highly-heterogeneous demand patterns; see details in \ref{simulation}. The results show that the estimation error in the generalized system cost is consistently less than 3\% for all tested scenarios. Therefore, we conclude that the proposed models are accurate enough for the planning-level task at hand.}
%
	
	\subsection{Features of optimal designs}
\begin{figure}[!htbp]
		\centering
		\begin{subfigure}{0.32\textwidth}
						\includegraphics[width = \textwidth]{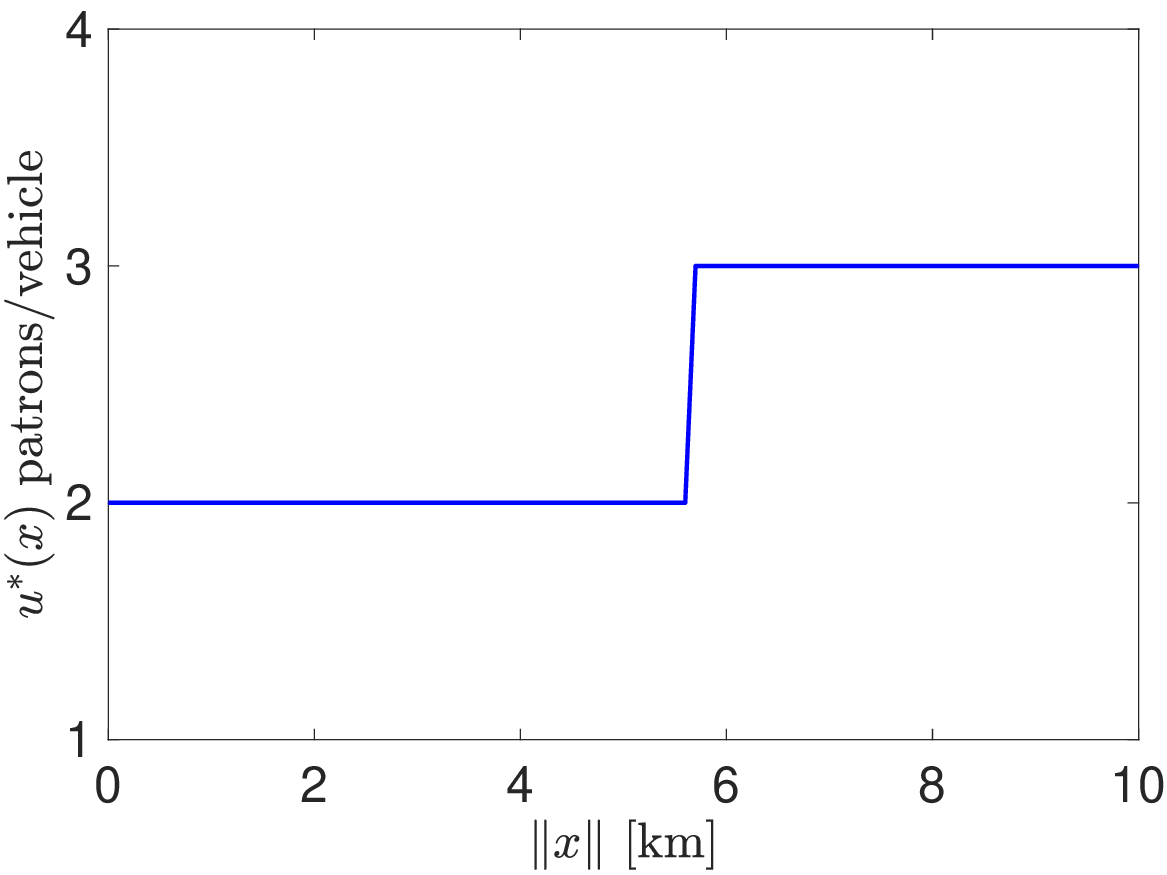}
			\caption{Changes of $ u^*(x) $ with $ \|x\| $}
		\end{subfigure}
		\begin{subfigure}{0.32\textwidth}
			\centering
						\includegraphics[width = \textwidth]{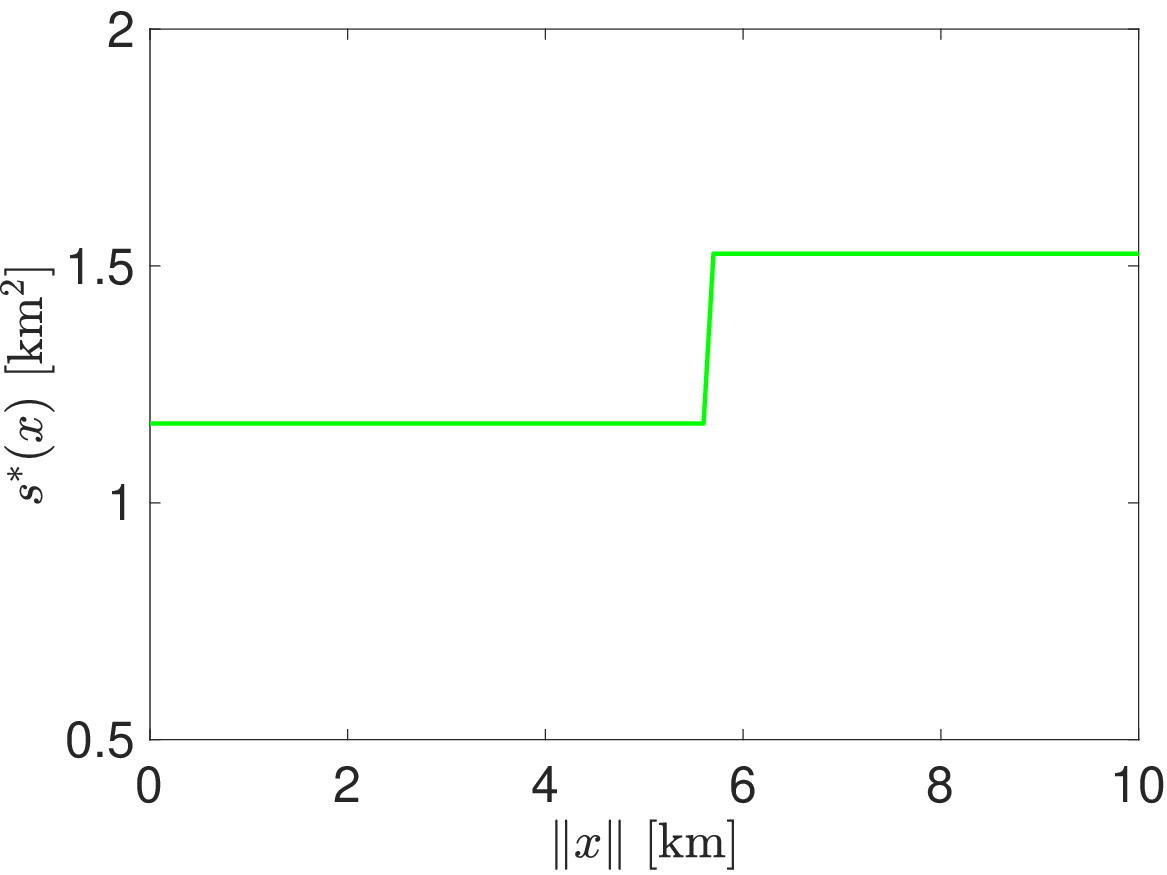}
			\caption{Changes of $ s^*(x) $ with $ \|x\| $}
		\end{subfigure}
		\begin{subfigure}{0.32\textwidth}
						\includegraphics[width = \textwidth]{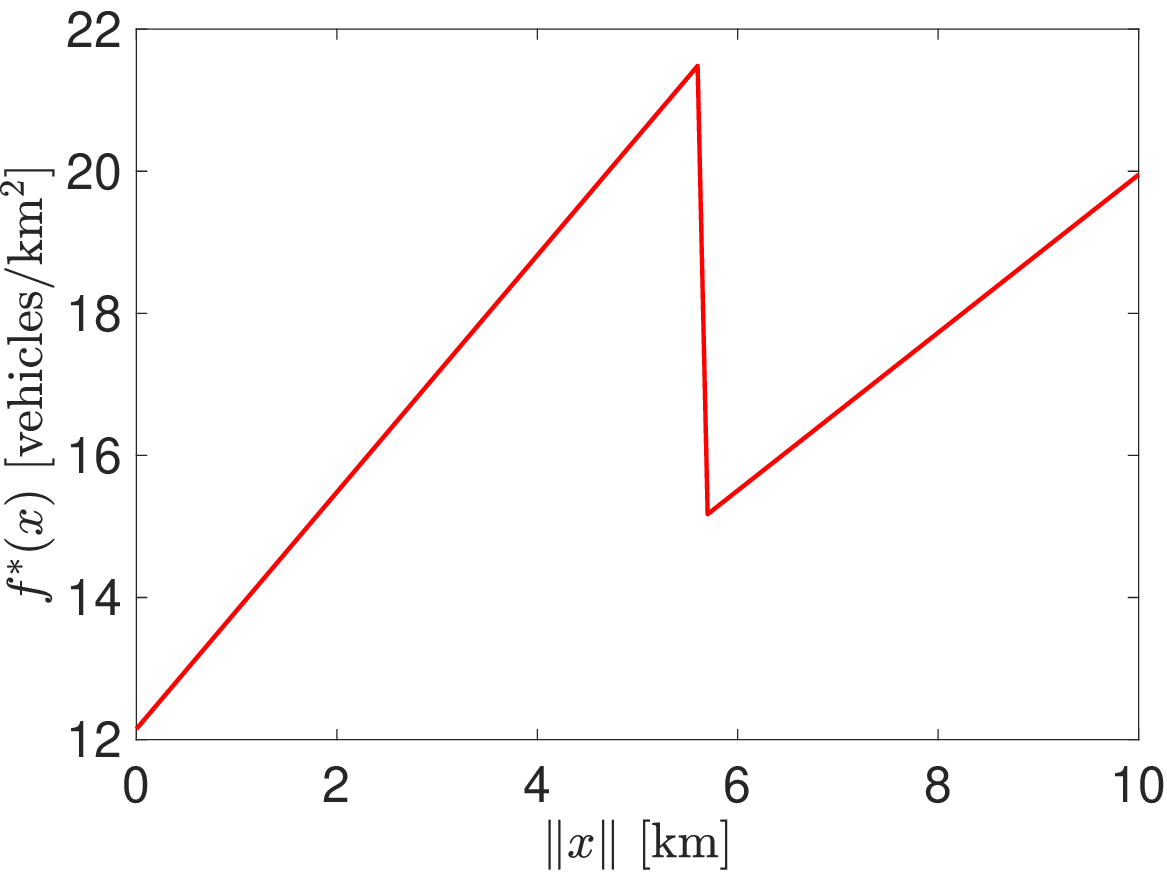}
			\caption{Changes of $ f^*(x) $ with $ \|x\| $}
		\end{subfigure}
		\caption{Optimized results under the uniform demand.}
				\label{optimized_results_vary}
	\end{figure}

	We first illustrate in Figure \ref{optimized_results_vary} how the optimized design variables vary with location under the \textit{spatially uniform} demand. 
	As seen in Figure \ref{optimized_results_vary}a, the optimized ride-pooling size $ u^*(x) $ increases with the distance $ \|x\| $. The result implies that transit services {operating at more distant locations should carry more passengers per vehicle to achieve greater transportation efficiency}. The optimized zone size $ s^*(x) $ in Figure \ref{optimized_results_vary}b exhibits the same trend as of $ u^*(x) $ under the uniform demand, as implied by Eq. (\ref{sol_s}). 
	Figure \ref{optimized_results_vary}c indicates that as the distance $ \|x\| $ grows, the optimized fleet density $ f^*(x) $ rises linearly but drops abruptly where $ u^*(x) $ increases. The result is not surprising because {an increased occupancy reduces the need for vehicles; see Eq. (\ref{sol_f})}.

	{The above trends are altered under spatially heterogeneous demands}. For instance, the distance-decaying demand may {result in non-piecewise linear $ s^*(x) $ and $ f^*(x) $ curves, and $ f^*(x) $ gradually decreasing with $ \|x\| $}. As for $ u^*(x) $, it tends to increase with the demand density; thus, {$ u^*(x) $ may diminish with $ \|x\| $ under a distance-decaying demand. The detailed results are omitted for simplicity.}
	
	These results confirm the need of developing a heterogeneous service design. The benefit of heterogeneous designs is discussed in the next subsection. 
	
	\subsection{Benefits of heterogeneous designs}
\begin{figure}[!htbp]
		\centering
        \begin{subfigure}{0.45\textwidth}		     \includegraphics[width = \textwidth]{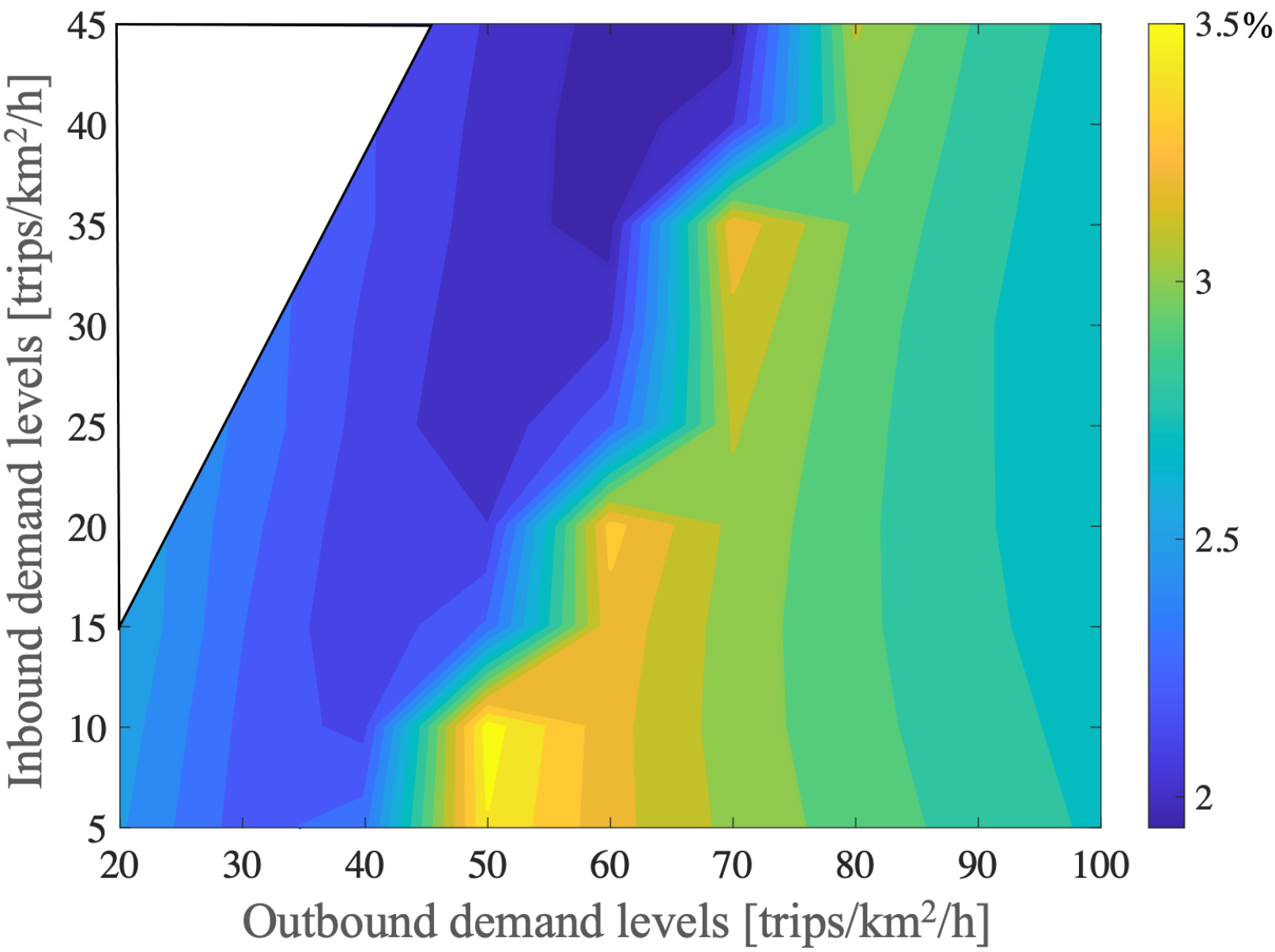}
			\caption{Cost savings against using the uniform $ u,s$, and $ f_0 $ under highly-heterogeneous demand} 
		\end{subfigure}
        \begin{subfigure}{0.45\textwidth}
			\includegraphics[width = \textwidth]{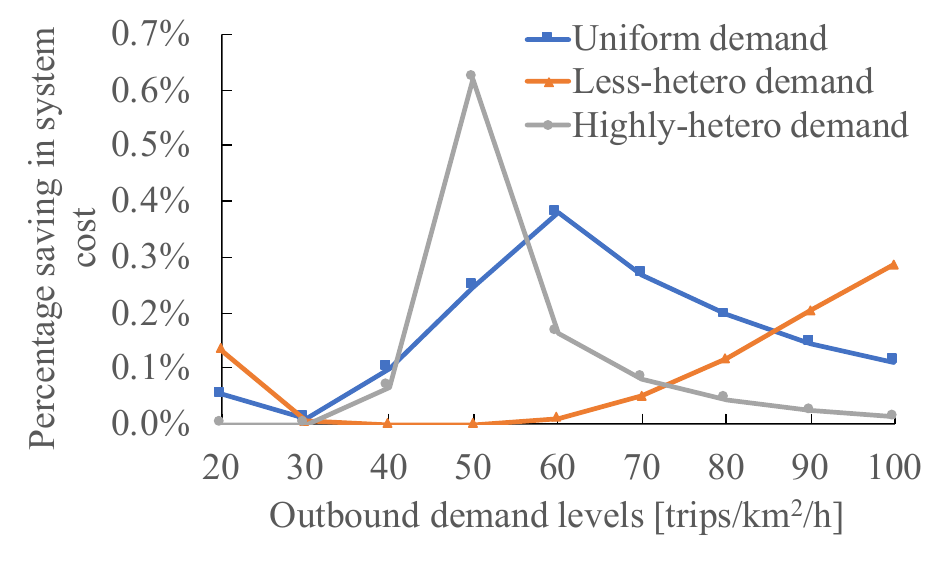}
			\caption{Cost savings against using a uniform $ u $}
		\end{subfigure}
  
		\begin{subfigure}{0.45\textwidth}
			\includegraphics[width = \textwidth]{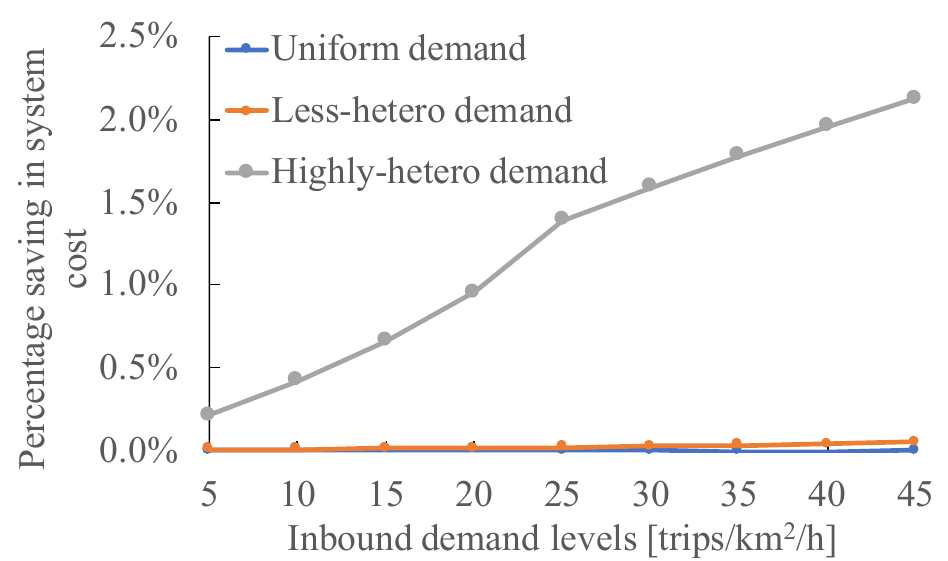}
			\caption{Cost savings against using a uniform $ s $}
		\end{subfigure}
		\begin{subfigure}{0.45\textwidth}
			\includegraphics[width = \textwidth]{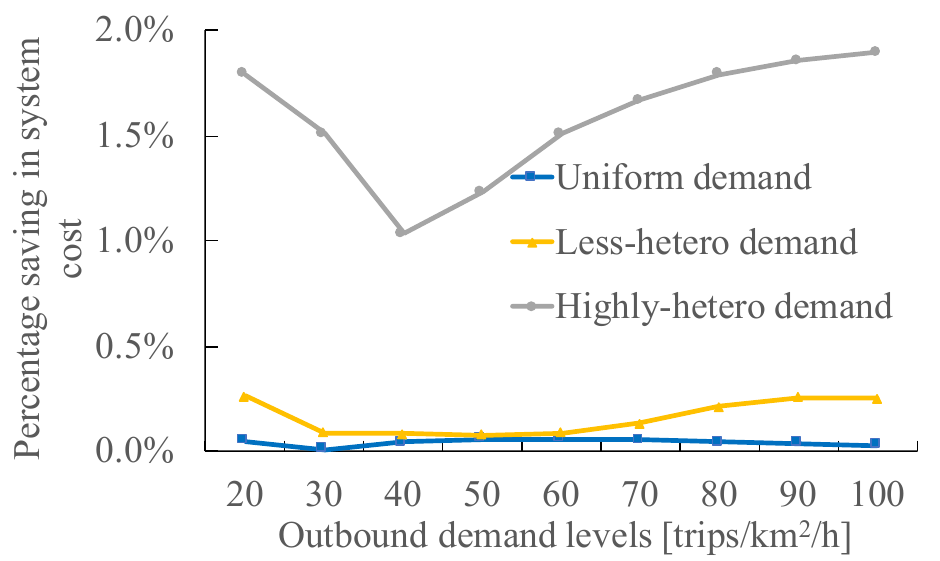}
			\caption{Cost savings against using a uniform $ f_0 $}
		\end{subfigure}
		
		\caption{Comparison against uniform designs. ({The horizontal axes of Figures \ref{vs_uniform_u}b, d represent the outbound demand, which has a critical influence on $u^*(x)$ and $f^*_0(x)$, as seen in Eqs. (\ref{z_u}, \ref{sol_f0}), while the horizontal axis of Figure \ref{vs_uniform_u}c is the inbound demand, which determines $s^*(x)$, as seen in Eq. (\ref{sol_s})}.) }
		\label{vs_uniform_u}
	\end{figure}

	We now examine the benefit of the spatially heterogeneous RP designs compared to the uniform designs. {Figure \ref{vs_uniform_u}a plots the percentage saving in system cost of the optimal heterogeneous design under the highly-heterogeneous demand, using the designs optimized for uniform $u, s$, and $f_0$ as the basis. The figure reveals that the cost saving is up to 3.5\%, which is attained when the inbound demand is small and the outbound demand is around 50 [trips/km$ ^2 $/h]. {This benefit accumulates with daily operation and could be considerable for the entire society. Rough estimates show that in a high-wage city with thousands of RP trips per day, the daily saving could amount to hundreds of thousands of dollars.} The result speaks to the importance of heterogeneous service designs.}
 
    {Comparisons are also made to the optimized designs where one of the three decision variables is spatially uniform while the other two are location-varying.} Figures \ref{vs_uniform_u}b--d show the percentage savings of system costs for these comparisons under three types of demand patterns (i.e., uniform, less-, and highly-heterogeneous demand). {The curves of three demand patterns do not follow the same pattern due to the combined impact of spatial heterogeneity in demand distribution and the integer-valued $u^*(x), x \in \mathbf{R}$.}
	
	{As expected, all the comparisons reveal positive cost savings. The greatest savings are observed under the highly-heterogeneous demand. The savings resulting from heterogenizing $u$ are less than 1\% (Figure \ref{vs_uniform_u}b), while heterogenizing $s$ and $f_0$ can produce up to 2\% savings under the highly-heterogeneous demand, as seen in Figures \ref{vs_uniform_u}c--d.}
	
	{We observe that the improvement in system performance is slight (yet consistently positive) under scenarios of uniform demand. This finding is favorable, suggesting that in practical cases where demand remains relatively invariable in space, uniform designs could be applied to streamline the operation of RP services without much compromising system performance.}

	\subsection{Benefits of the zoning strategy}
\begin{figure}[!htbp]
		\centering
		\begin{subfigure}{0.45\textwidth}
						\includegraphics[width = \textwidth]{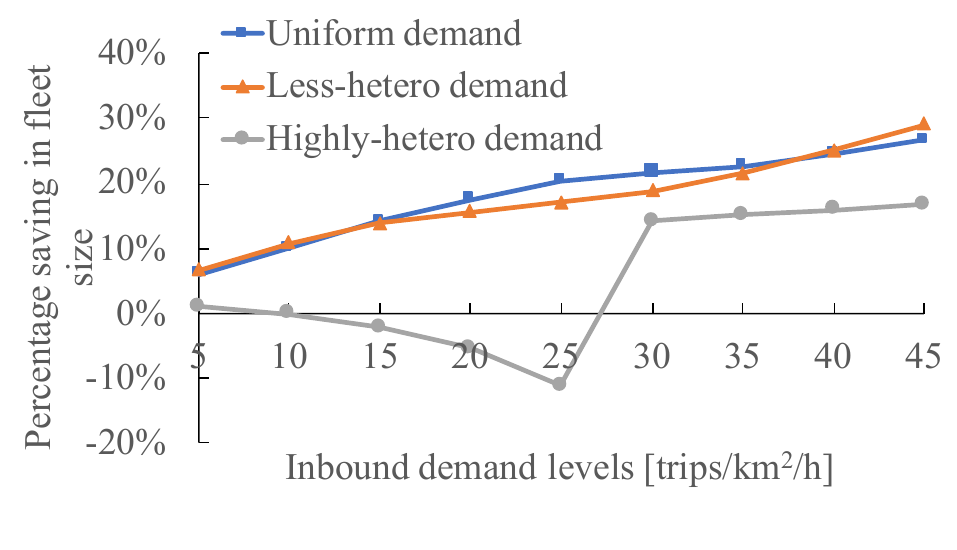}
			\caption{Changes in fleet size}
		\end{subfigure}
		\begin{subfigure}{0.45\textwidth}
						\includegraphics[width = \textwidth]{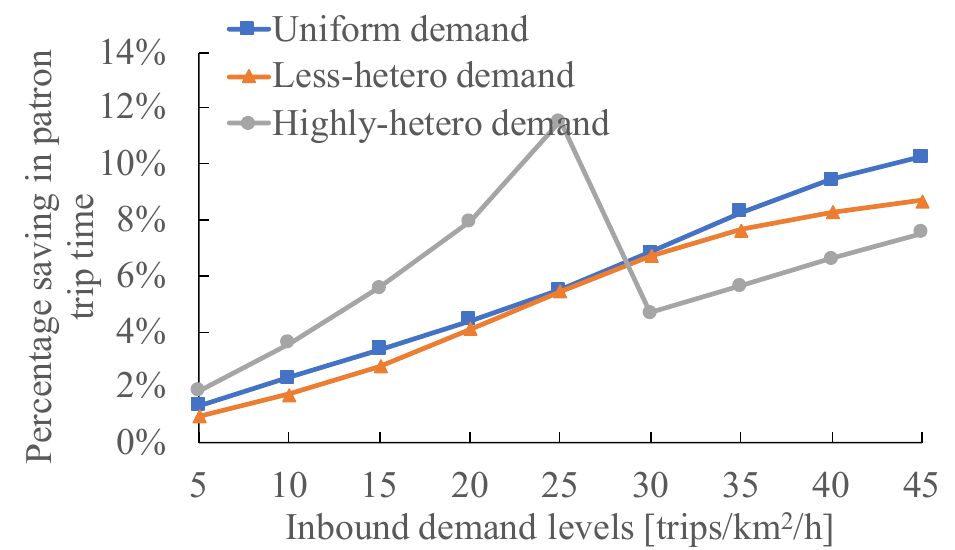}
			\caption{Changes in patron trip time}
		\end{subfigure}
		\begin{subfigure}{0.45\textwidth}
						\includegraphics[width = \textwidth]{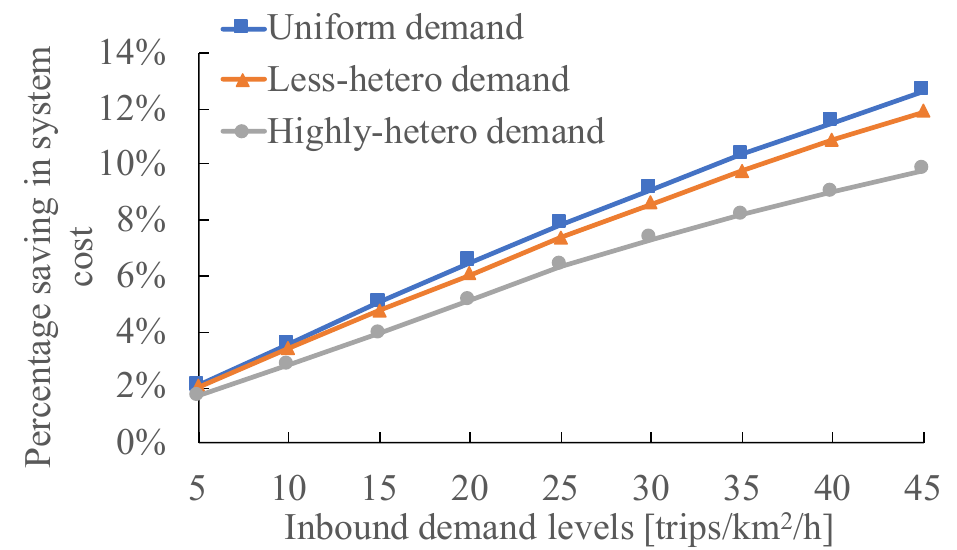}
			\caption{Savings in system cost}
		\end{subfigure}
		\caption{Comparisons against designs without zoning. }
				\label{vs_uniform_s}
	\end{figure}

	To demonstrate the effectiveness of the zoning strategy, we make comparisons to optimal designs without zoning, i.e., where $ s(x)=5 \times 5 = 25 \text{ km}^2,\forall x \in \mathbf{R} $. {The changes in the fleet size, patron trip time, and the system cost are depicted in Figure \ref{vs_uniform_s} with the inbound demand varying in $ [5, 45] $ [trips/km$^2$/h]. (We choose the inbound demand as the horizontal axis because it has a significant impact on the effect of zoning; see Section \ref{opt_rp}.) 
		As seen, the zoning strategy generally results in smaller fleet sizes and shorter patron trip times. The overall system cost saving is up to 12\%. The benefit generally grows with the inbound demand level and diminishes with the demand heterogeneity. This manifests the critical role played by the zoning strategy in RP designs.} 
	
	\subsection{{Comparison against two alternative strategies}} \label{sec_numerics}
	
	This subsection shows the advantage of the proposed RP-HD strategy over two alternatives: (i) RP-QD and (ii) FlexRT. They are described below.
	\begin{enumerate}[label=\roman*.]
		\item[(i)] RP-QD:
		{This strategy dispatches an RP vehicle immediately to pick up the first-come (outbound) request. The vehicle continues to receive new requests until the first patron is onboard or $ u $ requests are received} \citep{daganzo2019general,liu2021mobility}. Unlike our HD strategy, the QD strategy offers no waiting time for pooling requests but at the cost of sacrificing transportation efficiency due to the lower average vehicle occupancy.
		\item[(ii)] FlexRT:
		The FlexRT is a demand-responsive transit service adapted from the works of Professor Paul Schonfeld and his co-authors \citep[e.g.,][]{chen2022optimized,liu2020effects,kim2019optimal,kim2014integration,chang1991optimization}. FlexRT vehicles are dispatched at fixed headways. Each vehicle serves the requests received from a designated service zone during the headway prior to its dispatch. The major difference between FlexRT and the RP services is that a FlexRT vehicle needs to traverse an entire service zone to pick up the patrons, while an RP vehicle only accepts the closest requests to its location. 
		
	\end{enumerate}
	
	For a fair comparison, we extend the homogeneous design models in the literature \citep{daganzo2019general,liu2021mobility,chen2022optimized,kim2019optimal} to account for heterogeneous designs under spatially heterogeneous demands. The detailed models of RP-QD and FlexRT services are deferred to \ref{sec_QD} and \ref{sec_FlexRB}, respectively. Unless otherwise stated, the parameter values are kept the same as the RP-HD service. For simplicity, a constant ride-pooling size, $ u=2 $, is set in the RP-QD service according to \cite{daganzo2019general} and \cite{liu2021mobility}, because $u>2$ would greatly complicate the model.
	 
\begin{figure}[!h]
		\centering
		\begin{subfigure}{0.45\textwidth}
						\includegraphics[width = \textwidth]{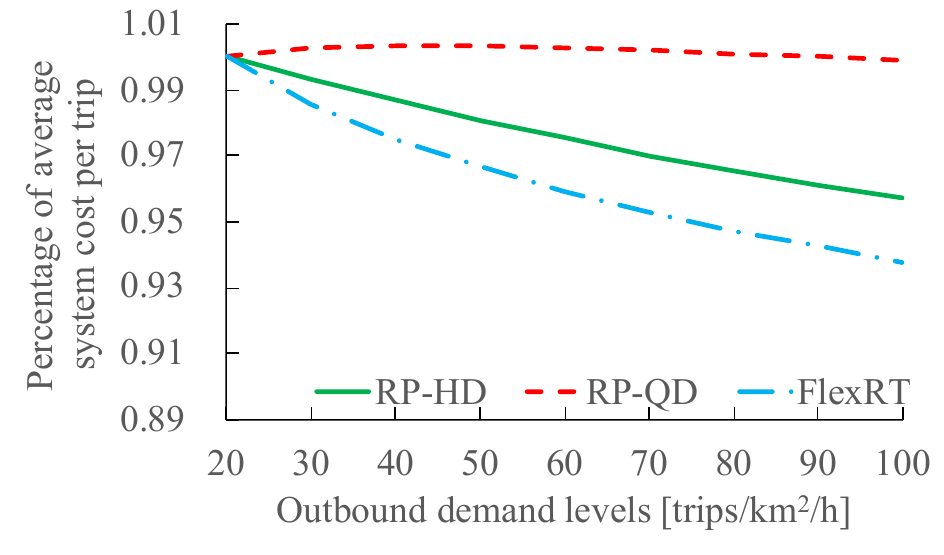}
			\caption{Uniform demand}
		\end{subfigure}
		\begin{subfigure}{0.45\textwidth}
			\centering
						\includegraphics[width = \textwidth]{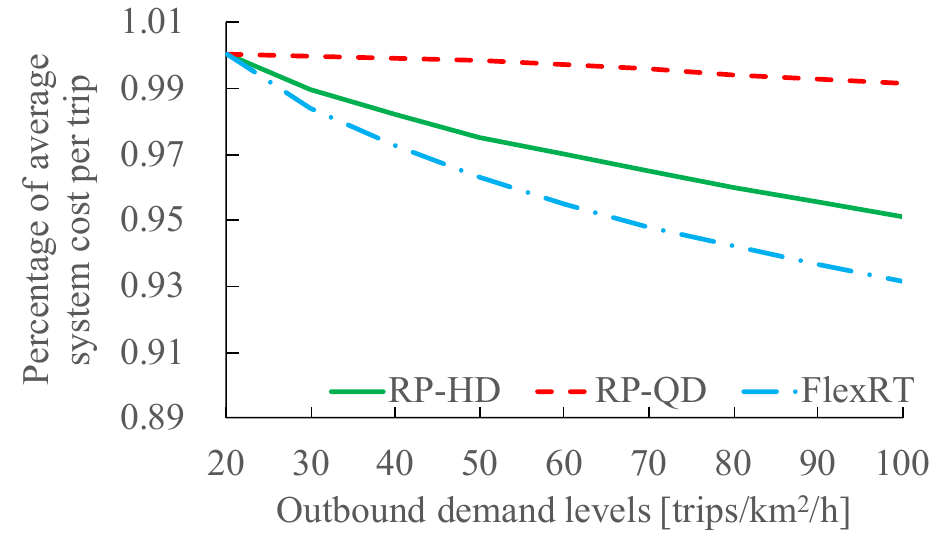}
			\caption{Less-heterogeneous demand}
		\end{subfigure}
		\begin{subfigure}{0.45\textwidth}
						\includegraphics[width = \textwidth]{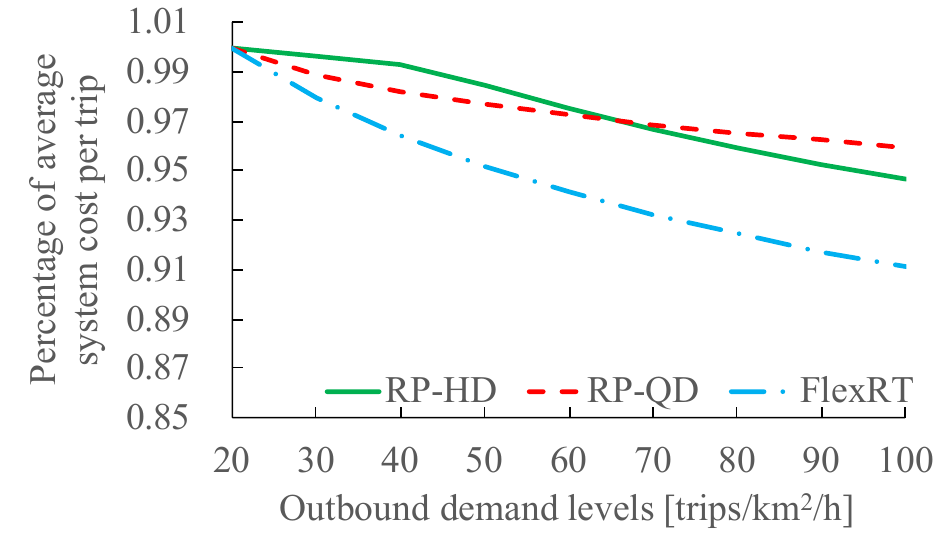}
			\caption{Highly-heterogeneous demand}
		\end{subfigure}
		
		\caption{Economies of scale for RP-HD, RP-QD, and FlexRT services.}
				\label{EOS_HD_QD}
	\end{figure}

	First, we find that RP-HD exhibits economies of scale (EOS). Figure \ref{EOS_HD_QD} plots the percentage of average system cost per trip, using $\bar{\lambda}_u = 20$ [trips/km$^2$/h] as the basis, against the outbound demand level, i.e., $\frac{\text{Average system cost per trip under } \bar{\lambda}_u \in (20, 100]}{\text{Average system cost per trip under } \bar{\lambda}_u = 20}\times 100\%$. Note that the average system cost per trip for the RP-HD service declines steadily from low to high demand levels under all types of demand patterns. This result is also implied by Eq. (\ref{z_u}); see that $\frac{z_u}{\lambda_u(x)}$ is a decreasing function of $\lambda_u(x)$. In contrast, RP-QD exhibits a modest EOS only under the highly-heterogeneous demand, {while FlexRT consistently yields a more significant EOS than RP-QD and RP-HD.}
	
	 The observation regarding RP-QD can be explained by its average vehicle occupancy, which is close to one (meaning little or no EOS) under the uniform and less-heterogeneous demands but nearly two under the highly-heterogeneous demand. It also implies that RP-QD's ride-pooling target (i.e., $ u=2 $ in this case) may have little impact on the system performance when the demand is low or less heterogeneous. In other words, RP-QD behaves like (e-hailing) taxis under these circumstances.
	
\begin{figure}[!h]
		\centering
		\begin{subfigure}{0.45\textwidth}
			\includegraphics[width = \textwidth]{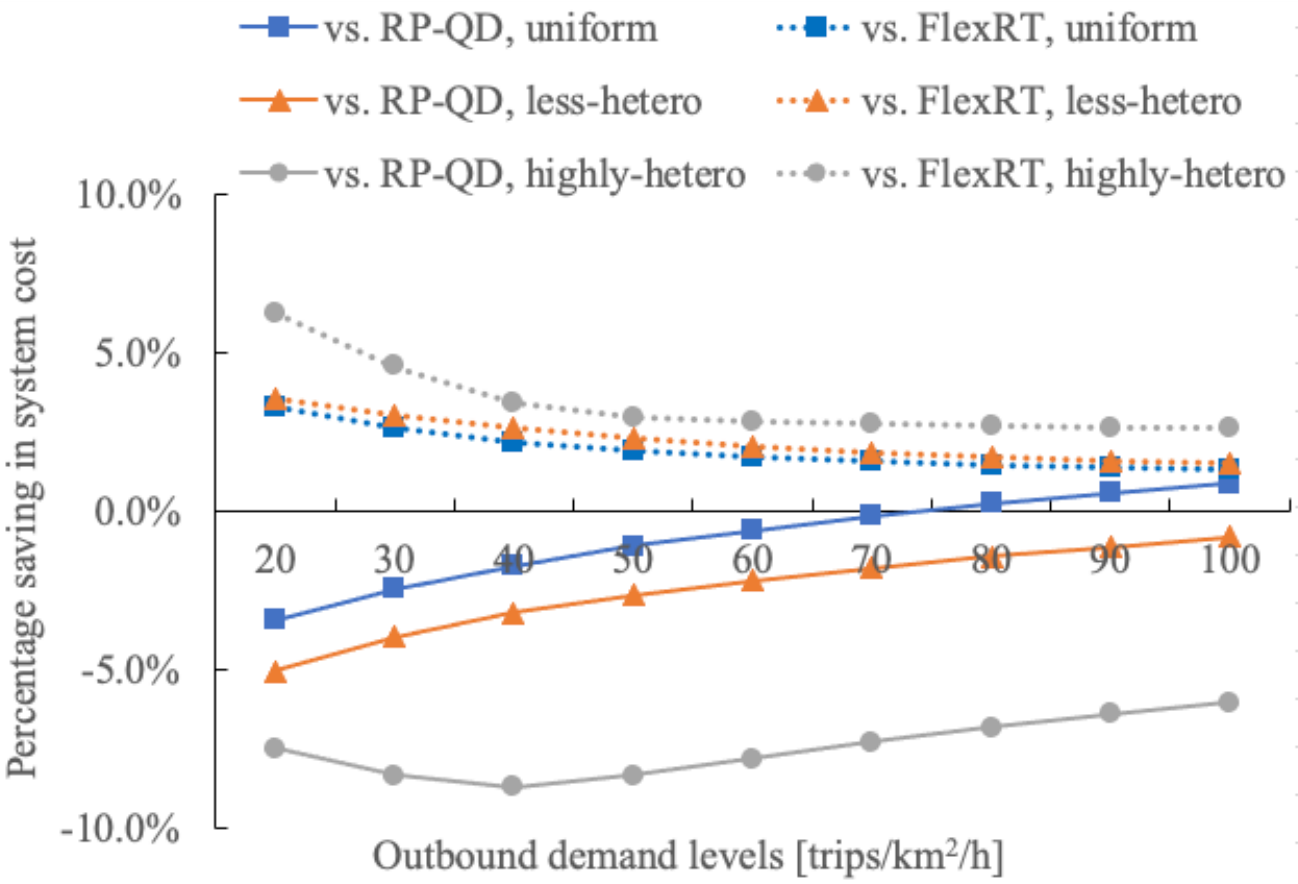}
			\caption{System cost savings under the basic setting}
		\end{subfigure}
		\begin{subfigure}{0.45\textwidth}
			\centering
			\includegraphics[width = \textwidth]{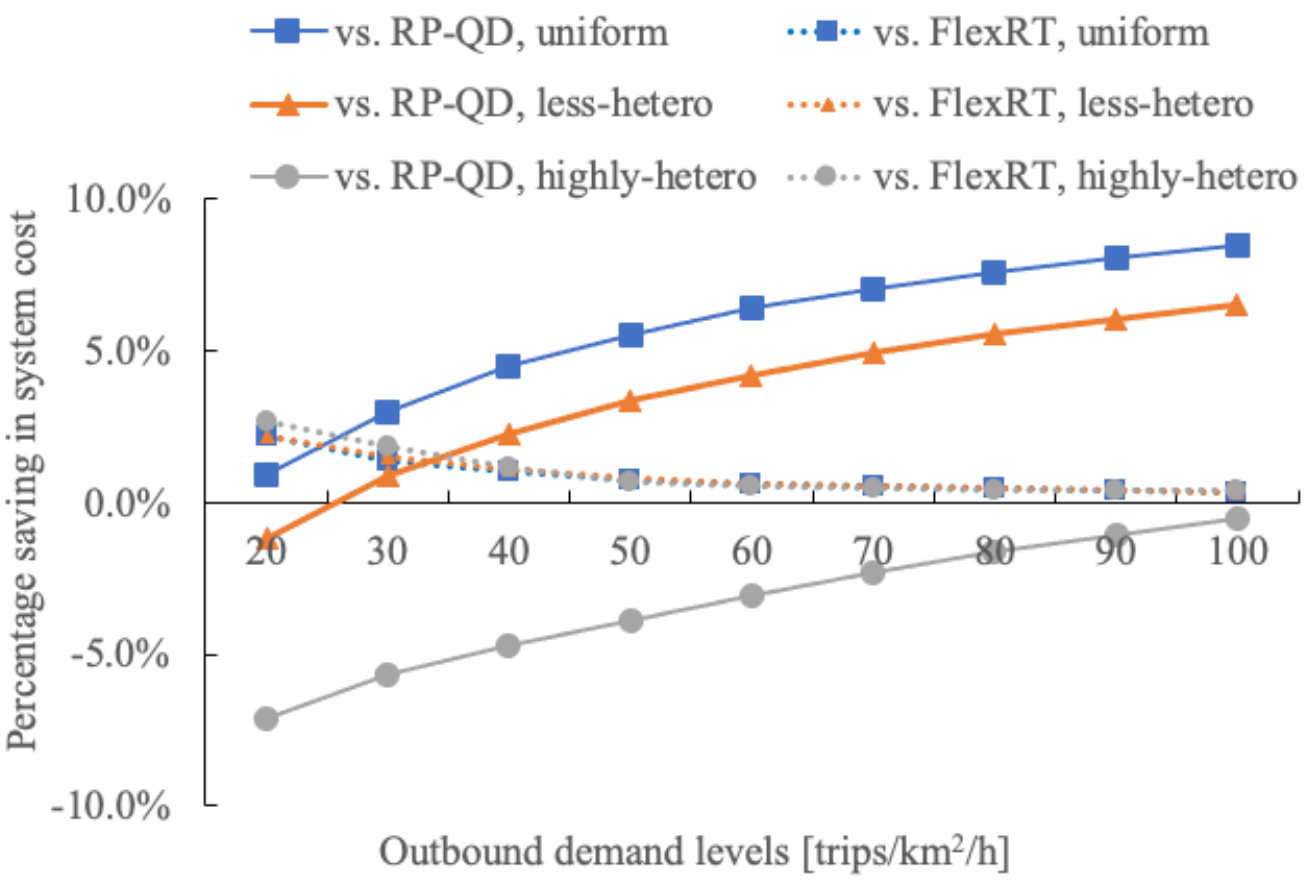}
			\caption{System cost savings with a higher unit operational cost}
		\end{subfigure}
		\begin{subfigure}{0.45\textwidth}
			\includegraphics[width = \textwidth]{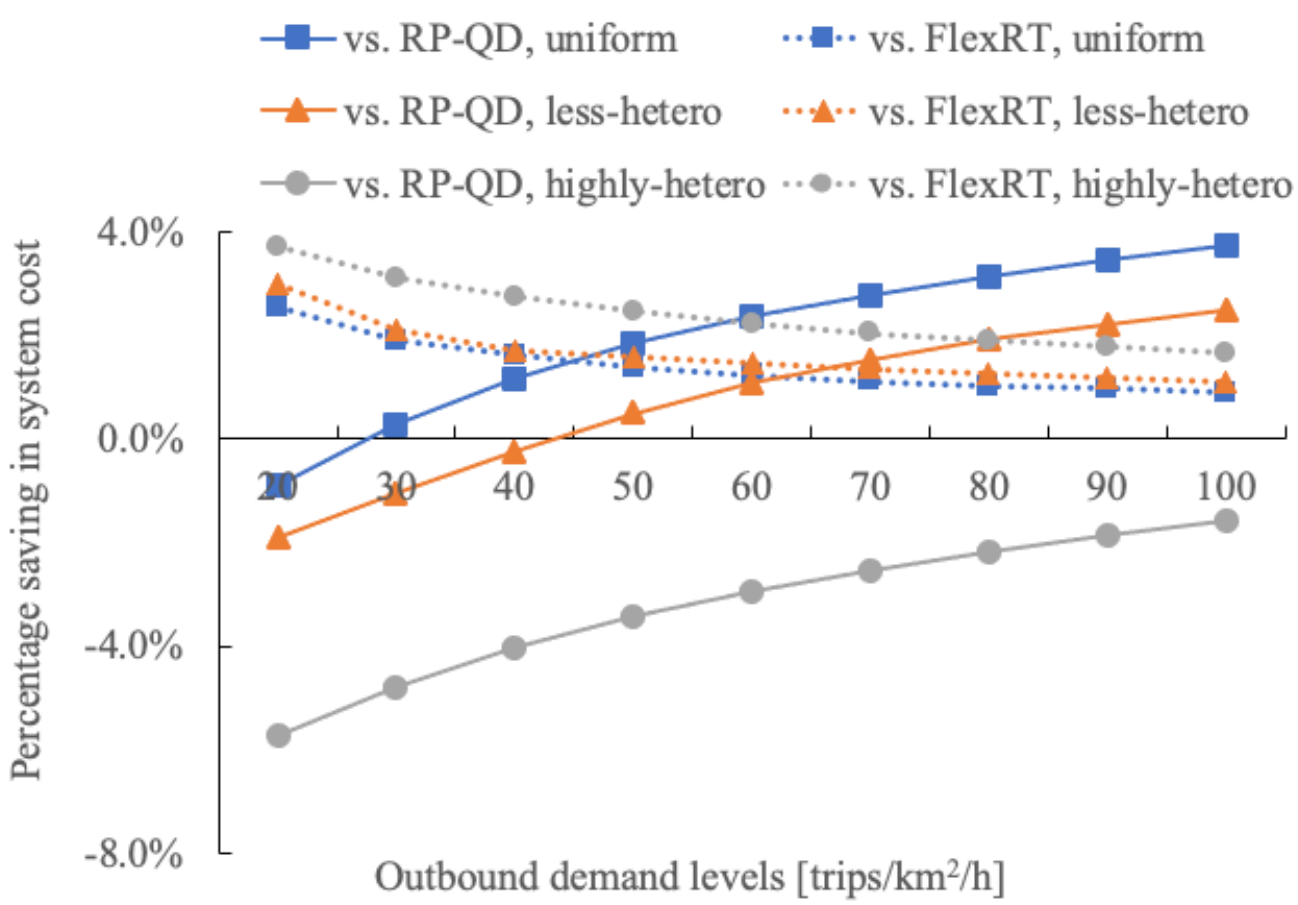}
			\caption{System cost savings with a longer line-haul distance}
		\end{subfigure}
		\begin{subfigure}{0.45\textwidth}
			\includegraphics[width = \textwidth]{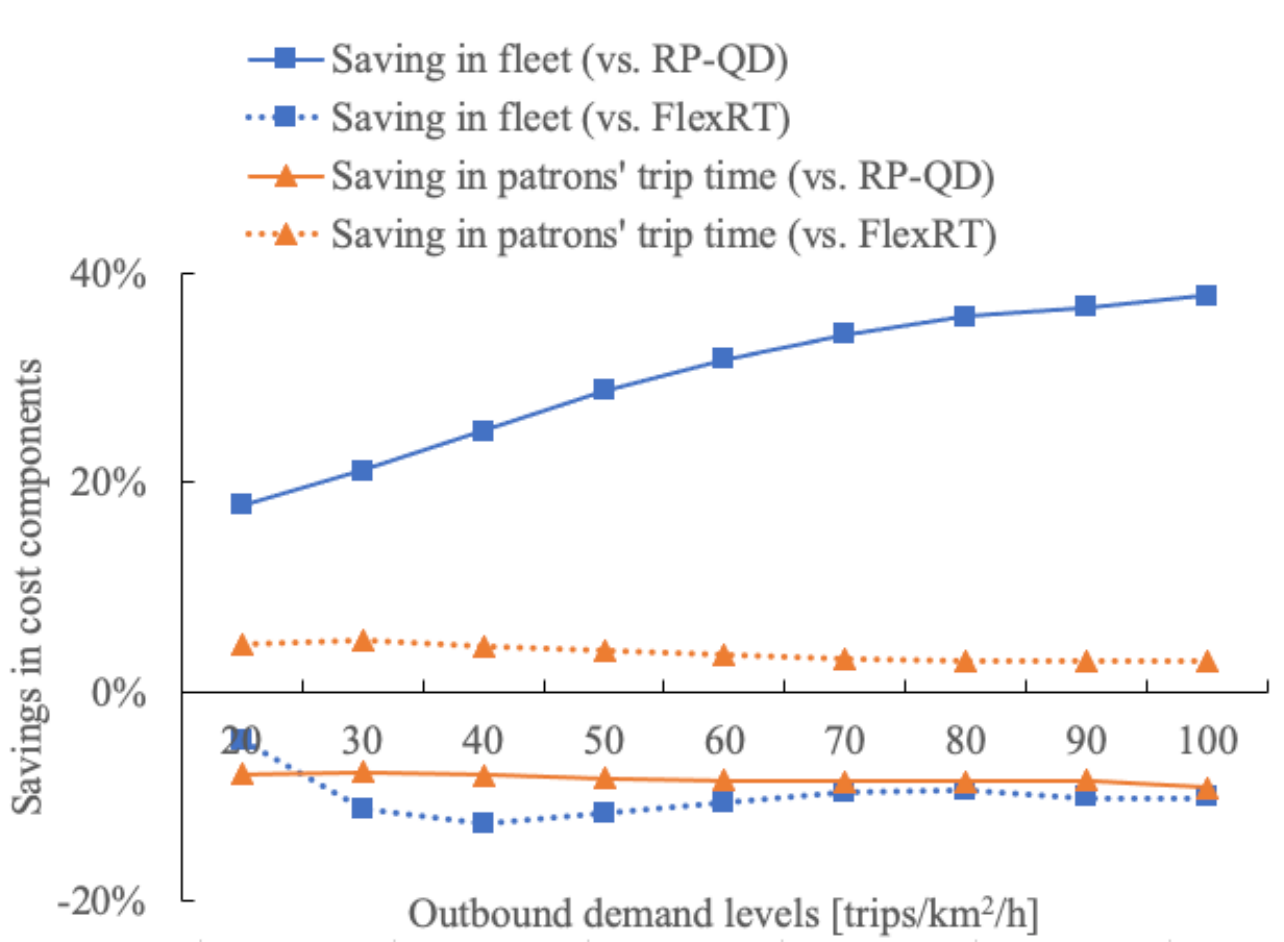}
			\caption{Savings in cost components under the uniform demand with the basic setting}
		\end{subfigure}
		
		\caption{Performance comparisons of RP-HD, RP-QD, and FlexRT.}
		\label{vs_counterparts}
	\end{figure}

	Next, we compare the performance of RP-HD to its counterparts. Figures \ref{vs_counterparts}a--c present the percentage system cost savings of RP-HD against the two alternative strategies under the three demand patterns with: (a) the default setting, (b) {a relatively higher unit operation cost ($\pi_{f}$ is doubled to $\$11.8 $ per vehicle hour; or equivalently, the VOT $ \mu $ is cut by half to $ 12.5 $ [\$/h])}, and (c) a longer line-haul distance ($ L=10 $ km). 
	
	{As compared to RP-QD,} RP-HD's benefit amplifies with higher demand levels, thanks to the EOS. RP-HD dominates in scenarios that call for higher transportation efficiency, i.e., when demand is more evenly distributed, the vehicle's unit operational cost is higher, or the line-haul distance is longer (which calls for a larger fleet). The system cost savings of RP-HD can exceed 8\% (see Figure \ref{vs_counterparts}b). This is mainly due to the reduction in fleet size; see Figure \ref{vs_counterparts}d for the savings in cost components.
	
	{As compared to FlexRT,} RP-HD achieves positive system cost savings across all tested scenarios. However, the trends in cost savings seem to be the opposite to those revealed in the comparison against RP-QD. Now RP-HD's benefit diminishes as the demand level rises, {thanks to the greater EOS of FlexRT (see Figure \ref{EOS_HD_QD}).} The benefit of RP-HD over FlexRT is larger when demand is more heterogeneous, the vehicles' unit operation cost is lower, or the line-haul distance is shorter. {RP-HD yields lower patron travel times than FlexRT at the cost of larger fleet sizes; see Figure \ref{vs_counterparts}d. Further tests show that FlexRT will outperform RP-HD when the fleet cost becomes higher or the patrons' value of time is lower.}	{In summary, the new RP system demonstrates its superiority across various scenarios, but its practical performance compared to its counterparts depends on local application conditions.}
	
	\section{Conclusions}
		This paper proposes analytical models for designing ride-pooling (RP) as on-demand feeder services connecting remote areas to transit terminals. 
		The models can furnish heterogeneous designs concerning location-dependent zone size, fleet density, and ride-pooling size to best accommodate the spatial heterogeneity arising from (i) non-uniform demand distributions and (ii) service locations with varying distances to the connected terminal. Thanks to the parsimony of the proposed models, closed-form formulas regarding the optimal zone size and fleet density are obtained, which unveil new insights. 
		
		We find that {the RP service enjoys flexibility in zone partitioning when the inbound demand is negligible (e.g., tidal commute demand in the morning peak hours); see point (i) in Section \ref{opt_rp} and \ref{sec_QD}. It is, however, not the case for flexible-route transit (FlexRT), where zone partitioning plays a more critical role in the optimal service design; see \ref{sec_FlexRB}. Furthermore, the spatial heterogeneity is rooted in the optimal service designs (see point (iii) in Section \ref{opt_rp}).}
		Numerical experiments show that RP with quick-dispatch (QD) strategy is inferior to the proposed hold-dispatch (HD) strategy when the demand is high and less heterogeneous and in the suburbs with high operation costs and long-distance line-haul connections (see Figure \ref{vs_counterparts}). 
		Moreover, RP-HD consistently outperforms FlexRT in all tested scenarios (see again Figure \ref{vs_counterparts}) except when the fleet cost is extremely high or the patrons' VOT is very low. 
		These findings are helpful for practitioners to choose the right form of on-demand mobility and to design and operate the services accordingly.  {The applications of the proposed models can be straightforward since all computations would be performed by computer programs in the background once the demand and parameter settings are input.}
		
		{The RP-HD strategy can be implemented in situations where either a single TNC dominates the market, or the competitive landscape among multiple TNCs is less aggressive. (Contrarily, the RP-QD strategy likely represents how TNCs operate under intensive market competition and must adopt a quick-dispatch strategy for fear of losing customers.) As third-party integrators like Baidu Map, Gaode Map, and Hellobike gain prominence in the Chinese market, they consolidate various ride-sourcing services from multiple providers on a single platform \citep{ZHOU202276}. These third-party integrators alleviate market competition, making it more feasible for TNCs to adopt the more efficient RP-HD strategy. Another method to promote RP-HD is offering differentiated RP services: while patrons of high VOT may prefer taxis or the RP-QD service with shorter waiting times, those of low VOT would choose the cheaper RP-HD option despite the potentially longer waiting.} 
	
	It is worth noting that the current study focuses on macro-level modeling and is therefore limited in operational details. The models are formulated deterministically based on the steady-state assumption and random fluctuations in the demand and supply sides are ignored. Over-saturated traffic is also neglected, which limits the application of the proposed models in high-demand areas with severe congestion \citep{AMIRGHOLY2016234}. Nevertheless, such a macro-level theory is still helpful to improve the understanding of a new mode of transit feeder service and to provide a tool for quick examination of the feasibility and cost-effectiveness. The idealized designs can also be fine-tuned to generate blueprints considering practical conditions, e.g., real street networks and local demographics. 
	
	The present work can be extended in several ways:
	First, the optimized solutions (as functions over space) are not yet ready for implementation. They must be transformed into real designs with specific partitions of zones in which vehicles can be deployed and operated. {Methods of space partitioning \citep{ouyang2006discretization} exploiting Voronoi diagram generation techniques are under exploration.}
	Second, the type and size of RP vehicles are treated as given and identical across the region, which is not necessarily so in practice. Considering the spatiotemporal variations in demand, a mixed fleet may be desirable, with larger-sized vehicles providing more efficient transport at peak times or in densely-populated zones and smaller vehicles operating during off-peak periods or in low-demand zones. Optimization can be performed to find the best composition of a mixed fleet and how they are assigned to various zones.
	Third, the proposed models of RP services can be embedded into joint optimization with transit network designs. In this regard, it is interesting to examine the effect of schedule coordination between the RP feeder (for inbound trips) and the trunk transit. Additional considerations can be given to operational topics of interest, such as pricing and fare-splitting strategies under elastic demand to achieve a system-optimal demand share among different modes. {Patrons' preferred time windows can be embedded in RP vehicle routing to facilitate synchronized transfers at the transit terminal}. {The impacts of traffic congestion and occasional oversaturation due to demand surges are also worth exploring.}

	\section*{Acknowledgments}
	The research was supported by the Sichuan Provincial Science and Technology Innovation Cooperation Funds (No. 2020YFH0038) and the National Natural Science Foundation of China (72288101; 72091513). {We appreciate the editor and three anonymous reviewers for their valuable comments, which greatly helped improve this paper.} 
	
	\appendix
	\setcounter{table}{0}
	\setcounter{figure}{0}
    \section{Notation{s}} \label{notation}
	The notations used in this paper are summarized in Table \ref{tab_notations}.
	\begin{table}[h] 
		\centering
		\caption{List of notations.}
		\resizebox{\textwidth}{!}{\begin{tabular} {| l | l | l |} 
			\hline 
			\textbf{Decision variables} & \textbf{Description} & \textbf{Unit}\\
			\hline 
			$ f(x),f_0(x) $ & {Densities of total and available vehicles at $x$}  & vehicle/km$ ^2 $\\
			\hline
			$ s(x) $& Zone size {at $x$} & km$ ^2 $/zone \\
			\hline
			$ u(x) $ & Ride-pooling size {at $x$} & patrons/vehicle \\
			\hline 
			\multicolumn{3}{|l|}{\textbf{Other variables and parameters}} \\
			\hline
			\makecell[l]{$ a_{ij}(x), p_{ij}(x)$ \\ $d_{u0}(x), d_{v0}(x) $}& \makecell[l]{Flow rates of assignment, collection, and delivery of outbound and inbound patrons at $x$ \\ Subscripts $i,u,v$ mean the number of onboard patrons and $j,0$ the number of assigned requests} & vehicles/h\\
			\hline
			$ C $& Vehicle capacity & patrons/vehicle\\
			\hline
			$ F $& Total fleet size & \\
			\hline
			{$ k $} & {The coefficient in the TSP tour formula} & \\
			\hline
			$ L $ & Line-haul distance & km \\
			\hline
			$ n_{ij}(x) $ & Number of vehicles in {the state of $ i $ patrons on-board and $ j $ requests to be picked up at $x$} & \\
			\hline
			$ t_p(x) $& Average travel time for picking up (denoted by subscript $p$) one patron {at $x$} & hour/patron \\
			\hline
			$ t^w_u(x), t^r_u(x),t^d_u(x) $& \makecell[l]{Average trip time components of outbound (denoted by subscript $u$) patrons {at $x$} \\ Superscripts $w,r,d$ indicate waiting time at home, time for picking up, and line-haul time}& hour/patron \\
			\hline
			$ t^w_v(x), t^s_v(x),t^r_v(x)$ & \makecell[l]{Average trip time components of inbound (denoted by subscript $v$) patrons {at $x$} \\ Superscripts $w,s,r$ indicate terminal waiting time, line-haul time, and time for dropping off}& hour/patron\\
			\hline
			$ U_a $& Total cost of the operator & \$ \\
			\hline
			$ U_u, U_v $ & {Total time of outbound ($u$) and inbound ($v$) patrons} & hour \\
			\hline
			$ v(x) $ & {Average number of inbound patrons per vehicle at $x$} & patrons/vehicle \\
			\hline
			$ V, V' $& {Cruise and commercial vehicle speeds} & km/h\\
			\hline
			$ Z $& Total generalized system cost & hour \\
			\hline
			$ \lambda_u(x),  \lambda_v(x) $ & {Demand densities of outbound ($u$) and inbound ($v$) patrons at $x$}& trips/km$ ^2 $/h \\
			\hline
			$ \ell_u(x), \ell_v(x) $ &	{Expected lengths of outbound ($u$) and inbound ($v$) TSP tours at $x$} &	km \\
			\hline
			$ \mu $& Patrons' average value of time & \$/hour\\
			\hline
			$ \pi_{f} $& Unit operational cost related to fleet $f$ & \$/vehicle-hour \\
			\hline
		\end{tabular}} \label{tab_notations}
	\end{table}

	
	\section{Simulation experiments}	\label{simulation}
	\setcounter{table}{0}
	{We conduct the simulation experiments in two steps using the basic settings provided in Section \ref{set-up}. Firstly, we translate the optimized results (e.g., $s^*(x),x\in\mathbf{R}$) into specific designs using a discretization recipe, which partitions the service region into rectangle zones\footnote{{The main ingredient of space partitioning is locating the zones to minimize $ \sum_{i=1}^{I}\vert \int_{ x \in \mathbf{R}^{(i)} } \frac{1}{s^*(x)}dx - 1 \rvert, \text{ where }  I \equiv \int_{ x \in \mathbf{R} } \frac{1}{s^*(x)}dx$ means the total number of service zones and $\mathbf{R}^{(i)}$ the domain of the $i^\text{th}$ zone. Detailed steps of the discretization recipe can be found in \cite{Mayan2022}. Note that this is a simplified discretization recipe aiming for roughly estimating the error. Determining the partition that best fits $s^*(x)$ is doable, but too complicated to be included in this paper. The complexity is due to the fact that $x$ is a two-dimensional coordinate. (If $x$ is one-dimensional, e.g., within a density function along a line, a simple method of discretization has been developed; see \cite{mei2021planning}.) This more accurate recipe will be explored in future research (and will theoretically produce a lower error).}} and determines accordingly the fleet size and ride-pooling size for each zone.}

	{Secondly, we simulate RP-HD schemes in the SUMO environment, which is an open-source traffic simulation platform \citep{SUMO2018}.\footnote{{The source files of the SUMO simulations are available upon request.}} We then average the recorded system metrics over 500 simulations. Table \ref{Estimation errors} summarizes the comparison results against CA models, of which the estimation errors are gauged by $ \left| \frac{\text{Approximated value} - \text{Simulated value}}{\text{Simulated value}} \right| \times 100\%$. As seen, the errors in generalized system cost do not exceed 3\% for the three demand patterns. Slightly higher errors are observed in patrons' trip time, but they are still within 4\%. }
		\begin{table}[h!]
			\centering
			\caption{Estimation errors of CA models.}
			\label{Estimation errors}
			\small
			\begin{tabular}{|c|c|c|c|}
				\hline
				\textbf{Demand patterns} & \textbf{Generalized system cost} & \textbf{Patrons' trip time} & \textbf{Operator cost}\\
				\hline
				Uniform demand & 1.91\% & 2.23\% & 0.12\% \\
				\hline
				Less-heterogeneous demand & 2.86\% & 3.32\% & 0.26\% \\
				\hline
				Highly-heterogeneous demand & 2.52\% & 2.93\% & 0.18\% \\
				\hline
			\end{tabular}
		\end{table}

		Note that we indeed have found higher approximation errors in scenarios when zone sizes are \textit{deliberately set} to be very large (e.g., $>10$ km$^2$) under certain demand patterns. Fortunately, these cases did not happen with optimized zone sizes $s^*(x)$, which are constrained from becoming excessively large in conditions of $\lambda_v(x) < \lambda_u(x)$; see Eq. (\ref{sol_s}). 
	
	\section{Modeling ride-pooling with quick-dispatch strategy (RP-QD)} \label{sec_QD} 
\begin{figure}[!h]
		\centering
		\includegraphics[width=0.4\linewidth]{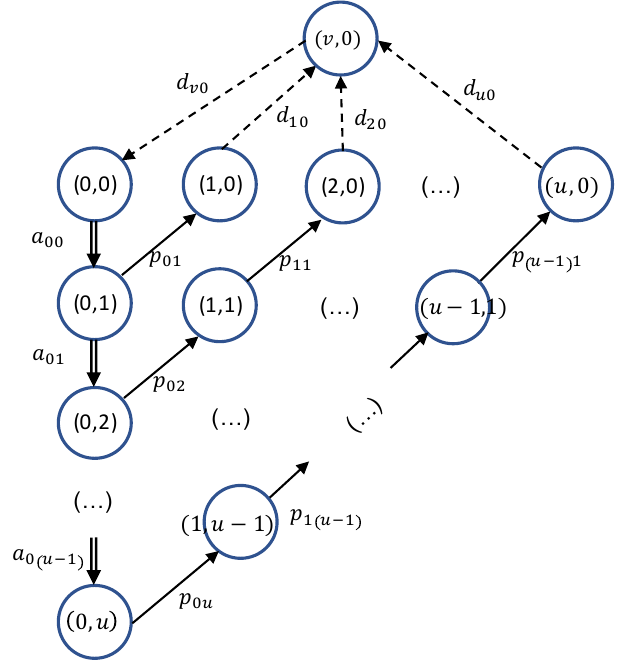}
		\caption{Workload transition network for vehicles in RP-QD.}
		\label{fig_QD_workload}
	\end{figure}

	The workload transition of vehicles under RP-QD {is illustrated} in Figure \ref{fig_QD_workload}.
	
	The flow rates between nodes in Figure \ref{fig_QD_workload} satisfy the following equations {in} steady states: 
	\begin{subequations}
		\begin{align}
			& a_{0j}(x)=\lambda_{u}(x)s(x)\frac{n_{0j}(x)}{\sum_{{k}=0}^{u(x)-1}n_{{0k}}(x)}=\lambda_{u}(x)\frac{n_{0j}(x)}{f_{0}(x)}, {j=0,1,...,u(x)-1,}\\
			& p_{\left(j-1\right)1} (x)=d_{j0}(x),j=1,2,...,u(x), \\
			& a_{0j}(x)=a_{0\left(j+1\right)}(x)+p_{0\left(j+1\right)}(x), j=0,...,u(x)-2, \\
			& a_{0\left(u-1\right)}(x)=p_{0u}(x)=d_{u0}(x),\\
			& a_{00}(x)=\sum_{i=1}^{u}d_{i0}(x)=d_{v0}(x)=\frac{\lambda_{v}(x)s(x)}{v(x)},\\
			& p_{0j}(x)=p_{1\left(j-1\right)}(x)=...=p_{\left(j-1\right)1}(x), j\in\left\{ 2,...,u(x)\right\}.
		\end{align}
	\end{subequations}

	We also know that 
	\begin{subequations}
		\begin{align}
			& \sum_{j=0}^{u(x)-1} {n_{0j}(x)} = f_0(x)s(x), \\
			& n_{ij}(x) = p_{ij}(x) t'_p(x), i\in\left\{ 0,1,...,u(x)-1\right\} ,j\in\left\{ 1,2,...,u(x)-i\right\} ,  \label{TSP_k'} \\
			& n_{i0}(x) = d_{i0}(x) t_{u}^{d}(x), i\in\left\{ 1,2,...,u(x)\right\} , \\
			& n_{v0}(x) = d_{v0}(x) t_{v}^{d}(x),
		\end{align}
	\end{subequations}
	where $ t'_p(x) $ in (\ref{TSP_k'}) is the average time {for a vehicle} to pick up one patron under RP-QD routing based on the proximity (rather than {following an optimal TSP tour }in RP-HD), and it can be estimated as follows, according to \cite{daganzo2019general}, 
	\begin{align}
		t'_p(x) =  \frac{ k'}{V'\sqrt{f_{0}(x)}},
	\end{align}
	where the dimensionless coefficient $ k' $ takes 0.63 in {the Manhattan metric} and 0.5 in the Euclidean metric.
	
	The $ t^d_u(x) $ and $ t^d_v(x) $ are the same as in Eqs. (\ref{t^d_u}, \ref{t^d_v}). 
	
	With the above relationships, $ \left\{n_{ij}(x)\right\} $ can be expressed as functions of $ f_0(x) $. The general expressions {are} complicated; for {simplicity, here} we present the results for the case of $ u(x)=2,\forall x\in{\mathbf{R}} $: 
	\begin{subequations}
		\begin{align}
			n_{00}(x)&=f_{0}(x)s(x)\frac{\left(\pi_{n}+1\right)}{\left(\pi_{n}+2\right)}, &&n_{01}(x)=\frac{f_{0}(x)s(x)}{\left(\pi_{n}+2\right)},\\
			n_{02}(x)&=\frac{f_{0}(x)s(x)}{\pi_{n}\left(\pi_{n}+2\right)},&&n_{10}(x)=\frac{\pi_{n}\lambda_{u}(x)s(x)}{\left(\pi_{n}+2\right)}\frac{\left(\|x\|+L\right)}{V},\\
			n_{11}(x)&=\frac{f_{0}(x)s(x)}{\pi_{n}\left(\pi_{n}+2\right)}, &&n_{20}(x)=\frac{\lambda_{u}(x)s(x)}{\left(\pi_{n}+2\right)}\frac{\left(\|x\|+L\right)}{V},
		\end{align}
		\begin{align}
			n_{v0}(x)&=\lambda_{u}(x)s(x)\frac{\left(\pi_{n}+1\right)}{\left(\pi_{n}+2\right)}\left(\ensuremath{\frac{\left(\|x\|+L\right)}{V}+\frac{k\sqrt{v(x)s(x)}}{V'}}\right), && 
		\end{align}
	\end{subequations}
	where $ \pi_{n} \equiv \frac{f_{0}(x)}{\lambda_{u}(x)}\frac{\sqrt{f_{0}(x)}V'}{k'} $ is a dimensionless variable dependent on $ x $ {and} $ f_0(x) $; and $ v(x)=\frac{\lambda_{v}(x)}{\lambda_{u}(x)}\frac{\left(\pi_{n}+2\right)}{\left(\pi_{n}+1\right)} $.
	
	We can then formulate the operator's cost and patrons' trip time as functions of $ f_0(x) $ {and $s(x)$} as follows. (The detailed derivations are omitted for the sake of brevity.)
	

	\begin{align}
		U_{a}=\pi_{f}\int_{x\in{\mathbf{R}}}\left[f_{0}(x)+\frac{2f_{0}(x)}{\pi_{n}\left(\pi_{n}+2\right)}+\lambda_{u}(x)\frac{\left(\pi_{n}+1\right)}{\left(\pi_{n}+2\right)}\left(\frac{2\left(\|x\|+L\right)}{V}+\sqrt{\frac{\lambda_{v}(x)}{\lambda_{u}(x)}\frac{\left(\pi_{n}+2\right)}{\left(\pi_{n}+1\right)}}\frac{k\sqrt{s(x)}}{V'}\right)\right]dx.
	\end{align}

	
	\begin{align}
		U_u =\int_{x\in{\mathbf{R}}}\left[\left(\frac{f_{0}(x)\left(\pi_{n}+4\right)}{\pi_{n}\left(\pi_{n}+2\right)}\right)+\left(\frac{\lambda_{u}(x)\left(\|x\|+L\right)}{V}\right)\right]dx.
	\end{align}
	
	\begin{align}
		U_{v}&=\int_{x\in{\mathbf{R}}}\left[\frac{1}{2s(x)}\frac{\lambda_{v}(x)}{\lambda_{u}(x)}\frac{\left(\pi_{n}+2\right)}{\left(\pi_{n}+1\right)}+\lambda_{v}(x)\left(\frac{\left(\|x\|+L\right)}{V}+\sqrt{\frac{\lambda_{v}(x)}{\lambda_{u}(x)}\frac{\left(\pi_{n}+2\right)}{\left(\pi_{n}+1\right)}}\frac{k\sqrt{s(x)}}{2V'}\right)\right]dx.
	\end{align}
	
	The corresponding optimal design problem is thus{:}
	\begin{align}
		\min_{f_0(x), s(x)} Z = \frac{U_a}{\mu} + U_u + U_v, \text{ subject to: } f_0(x),s(x) >0, \forall x \in {\mathbf{R}}.
	\end{align}
	
	Although there are no closed-form solutions {even for the simple case of $u(x)=2$}, optimal numerical results for each $ x \in {\mathbf{R}} $ can be found using off-the-shelf commercial tools, e.g., ``fmincon'' in MATLAB. 
	
	Note that the {optimal zoning} of RP-QD is again a result of the inbound demand; see in the above equations how terms related to $ s(x) $ vanish if $ \lambda_{v}(x) \approx 0, \forall x \in {\mathbf{R}} $.
	
	\section{Modeling flexible-route transit (FlexRT)} \label{sec_FlexRB}
	{For the FlexRT service, vehicles are dispatched at fixed headways to serve all the requests received in a headway. Before a vehicle is dispatched, an optimal TSP route is generated to match the received requests. We denote $ s(x) $ and $ h(x) $ as the spatially heterogeneous zone size and headway, respectively.}
	
	{We adopt the same assumptions made in the RP service:} No new requests will be accepted by {a vehicle that has completed ride matching and had its optimal route} generated, and vehicles execute the delivery and pickup tasks separately. 
	
	Under the above assumptions, the trip time of a FlexRT patron is composed of {four} parts: (i) the average waiting time {before the vehicle is dispatched (i.e., when the vehicle's requests are being received),} $ \frac{h(x)}{2}  $, identical for outbound and inbound patrons\footnote{{We assume that the trunk-line transit service frequency is high and there is no schedule coordination between the trunk-line and FlexRT services.}}{;} (ii) {outbound patrons' average waiting time for pick-ups, $ k\frac{\sqrt{u(x)s(x)}}{2V'} $; (iii) the average in-vehicle travel time in the local tour, $ k\frac{\sqrt{u(x)s(x)}}{2V'} $ for outbound patrons and $ k\frac{\sqrt{v(x)s(x)}}{2V'}  $ for inbound patrons; and (iv)} the line-haul travel time, $\frac{\|x\| + L }{V}$, identical for both types of patrons. 
	
	The heterogeneous {service} design model for FlexRT is formulated as the following optimization problem concerning $ s(x), h(x)$:
	\begin{linenomath*}
		\begin{subequations} \label{FlexRT_prob}
			\begin{align} \label{appen_obj}
				\min_{s(x),h(x)}\int_{x\in{\mathbf{R}}}{z(x,s(x),h(x))dx}
			\end{align}
			subject to:
			\begin{align}
				& u(x) = \lambda_{u}(x)s(x)h(x), \\
				& v(x) = \lambda_{v}(x)s(x)h(x), \\
				&  \max \left\{u(x), v(x)\right\} \le C,\\
				& s(x), h(x) > 0, \forall x \in {\mathbf{R}},
			\end{align}
		\end{subequations}
	\end{linenomath*}
	where the integrand in (\ref{appen_obj}) is {(with $u(x) = \lambda_{u}(x)s(x)h(x)$ and $v(x) = \lambda_{v}(x)s(x)h(x)$ plugged in):} 
	\begin{linenomath*}
	\begin{align} \label{local_obj}
		z(x,s(x),h(x)) & =  \frac{\pi_{f}}{\mu s(x)h(x)}\left(k\frac{s(x)\left(\sqrt{\lambda_{u}(x)h(x)}+\sqrt{\lambda_{v}(x)h(x)}\right)}{V'}+2\frac{\|x\|+L}{V}\right) \notag \\
		& +\left(\frac{\lambda_{u}(x)+\lambda_{v}(x)}{2}h(x)+\frac{ks(x)}{V'}\left(\sqrt{\lambda_{u}^{3}(x)h(x)}+\frac{1}{2}\sqrt{\lambda_{v}^{3}(x)h(x)} \right)+\left(\lambda_{u}(x)+\lambda_{v}(x)\right)\frac{\|x\|+L}{V}\right).
	\end{align}
	\end{linenomath*}

	Note that (i) {(\ref{FlexRT_prob}) can be decomposed by each $ x \in \mathbf{R} $, and it is minimized when (\ref{local_obj}) is minimized for all $ x \in \mathbf{R} $} subject to the constraints; and (ii) (\ref{local_obj}) {and the constraints are posynomial functions concerning $ s(x) $ and $ h(x) $, meaning that there exists a unique global minimum} according to geometric programming \citep{boyd2004convex}. 
	
	{Some analytical} results can be found {from the first-order conditions by defining} $ S(x) \equiv s(x)h(x), H(x) \equiv  \sqrt{h(x)} ${. Then,} (\ref{local_obj}) can be rewritten as
	\begin{align}
		z(x,S(x),H(x))&=\frac{\pi_{f}}{\mu S(x)}\left(k\frac{S(x)\left(\sqrt{\lambda_{u}(x)}+\sqrt{\lambda_{v}(x)}\right)}{V'H(x)}+2\frac{\|x\|+L}{V}\right) \notag \\
		&+\left(\frac{\lambda_{u}(x)+\lambda_{v}(x)}{2}\left(H(x)\right)^{2}+\frac{kS(x)}{V'H(x)}\left(\sqrt{\lambda_{u}^{3}(x)}+\frac{1}{2}\sqrt{\lambda_{v}^{3}(x)}\right)+\left(\lambda_{u}(x)+\lambda_{v}(x)\right)\frac{\|x\|+L}{V}\right),
	\end{align}
	of which the first-order conditions ({with the boundary constraints combined}) yield
	\begin{subequations} \label{opt_cond_flexRT}
		\begin{align}
			S^{*}(x)&=\min \left\{ \sqrt{\frac{2H^{*}(x)\pi_{f}\left(\|x\|+L\right)}{\mu k\left(\sqrt{\lambda_{u}^{3}(x)}+\frac{1}{2}\sqrt{\lambda_{v}^{3}(x)}\right)}\frac{V'}{V}}, \frac{C}{\lambda_{u}(x)}\right\}, \label{sol_s_FlexRT} \\
			H^{*}(x)&=\left(k\frac{\pi_{f}\left(\sqrt{\lambda_{u}(x)}+\sqrt{\lambda_{v}(x)}\right)+\mu S^{*}(x)\left(\sqrt{\lambda_{u}^{3}(x)}+\frac{1}{2}\sqrt{\lambda_{v}^{3}(x)}\right)}{\mu V'\left(\lambda_{u}(x)+\lambda_{v}(x)\right)}\right)^{\frac{1}{3}}.
		\end{align}
	\end{subequations}
	
	
	According to (\ref{opt_cond_flexRT}), an iteration algorithm can be employed to efficiently find the optimal solutions of $ \langle S^*(x), H^*(x) \rangle $ and $ \langle s^*(x), h^*(x) \rangle${, $ \forall x \in \mathbf{R}$}.
	
	Note that {for the FlexRT service, the optimal zone sizes cannot be freely determined} even if $ \lambda_{v}(x) \rightarrow 0, \forall x \in {\mathbf{R}} ${. This is different from the case of RP-HD, where the zone sizes can be arbitrarily determined when $ \lambda_{v}(x)=0$ (see insight (i) in Section \ref{opt_rp}). Thus, optimal zoning is more important for FlexRT. To see why, note that in FlexRT each vehicle traverses an entire zone to serve the patrons and the tour length is dictated by the zone size. In contrast, recall in the RP services that multiple vehicles are spread out in a zone and each vehicle only serves the requests arising in its neighborhood.} 

	\bibliography{reference}
\end{document}